\documentclass[%
reprint,
superscriptaddress,
nofootinbib,
nobibnotes,
 amsmath,amssymb,
 aps,
longbibliography
]{revtex4-2}
\usepackage[english]{babel}
\usepackage{graphicx}
\usepackage{dcolumn}
\usepackage{bm}

\usepackage[dvipsnames]{xcolor}
\usepackage[colorlinks=true,citecolor=LimeGreen,linkcolor=blue, urlcolor = cyan]{hyperref} 

\usepackage{color}
\usepackage[english]{babel}

\newcommand{\Tr}{\mathop{\rm Tr}\nolimits}

\begin{document}

\title{Diffusive-hydrodynamic transition in the anomalous Hall effect}

\author{D.S. Zohrabyan}
\affiliation{Moscow Institute of Physics and Technology, 141701 Dolgoprudny, Russia}
\affiliation{L. D. Landau Institute for Theoretical Physics, 142432 Chernogolovka, Russia}
\author{M.M. Glazov}
\affiliation{Ioffe Institute, 194021 St. Petersburg, Russia }%


\begin{abstract} 
We have developed a theory of the anomalous Hall effect in two-dimensional electron gas in the case where the time of electron-electron collisions is much smaller than the transport relaxation time. The transition between the diffusion transport regime, when the momentum relaxation length of electrons is much smaller than the channel width, and the hydrodynamic regime, when the momentum relaxation length exceeds the channel width, has been traced. The contributions of the anomalous velocity, wave packet shifts, and asymmetric scattering to the anomalous Hall field and voltage have been calculated. It has been shown that the anomalous Hall voltage caused by the asymmetric scattering can have a nontrivial coordinate dependence and change its sign depending on the specific scattering mechanism.

\end{abstract}

\maketitle

\section{\label{sec:intro}Introduction}

In recent years, the studies of the spin and anomalous Hall effects have became topical~\cite{dyakonov_book,RevModPhys.82.1539}. In these phenomena the spin degrees of freedom of charge carriers and the spin-orbit interaction is clearly manifested. The spin Hall effect (SHE) consists of generating a spin flux perpendicular to the applied electric field~\cite{dyakonov71,dyakonov71a,kato04,wunderlich05}. In multi-valley semiconductors, such as atomically thin transition metal dichalcogenides, a valley Hall effect (VHE) is possible, where carriers in different valleys propagate in opposite directions~\cite{Xiao:2012cr,2020arXiv200405091G}. In the anomalous Hall effect, the spin flux is converted into electric current in presence of external magnetic field. This leads to anomalous contributions to the Hall voltage and Hall constant, unrelated to an action of the Lorentz force on the charge carriers, but proportional to the spin polarization of the system~\cite{Hall:1881aa}.

The microscopic mechanisms of the spin and anomalous Hall effects are closely related to each other and have been actively discussed in the literature over the last half a century~\cite{PhysRev.95.1154,SMIT1955877,Adams:1959aa,gy61,abakumov72,nozieresAHE,PhysRevB.75.045315,Sinitsyn_2007,Ado_2015,PhysRevLett.123.126603}. By now, it is firmly established that there are three main mechanisms for these effects in non-magnetic semiconductors: (i) asymmetric scattering on impurities or phonons (skew scattering), (ii) shifts of electron wave packets during scattering (side-jump) and (iii) anomalous velocity, induced by an external electric field~\cite{Sinitsyn_2007}. Since the nature of the anomalous velocity and the side-jump contribution is the same and these effects are caused by the action on the electron of either an external field or a field created by a static defect or phonon, then under the steady state conditions a part of the side-jump contribution compensates the contribution of the anomalous velocity~\cite{dyakonov_book,2020arXiv200405091G}, see also~\cite{belinicher82}.

Nowadays, the spin and anomalous Hall effects have been studied in detail for the diffusive regime of the electron transport. This is a typical situation for sufficiently large, macroscopic samples, whose geometric dimensions exceed both the electron mean free path $l$, caused by the scattering of electrons on impurities and phonons, and the spin relaxation length $l_s$. However, the development of nanotechnology and progress in materials science make it possible to create ultraclean structures with two-dimensional electron gas, where the mean free path exceeds the width of the conducting channel $l \gg w$. In this case, the electron transport turns out to be qualitatively different compared to the diffusive case. A striking example is the hydrodynamic flow of electrons, recently discovered in ultraclean electronic systems~\cite{PhysRevB.51.13389,Bandurin1055,Crossno:2016aa,PhysRevLett.117.166601,Moll1061,Krishna-Kumar:2017wn,Gusev:2018tg,PhysRevLett.128.136801}. In this regime, the mean free path with respect to the interparticle collisions $l_{ee}$ is small compared to the channel width. Accordingly, the loss of momentum of charge carriers occurs mainly at the scattering on the channel edges, while the interparticle collisions provide the viscosity of the electron liquid. This leads to a number of nontrivial effects in the transport and magnetotransport of electrons in ultraclean systems ~\cite{gurzhi63,Gurzhi_1968,PhysRevLett.106.256804,PhysRevB.92.165433,Levitov:2016aa,PhysRevLett.118.226601,PhysRevB.100.115401}, see also reviews~\cite {Narozhny:2017vc,Narozhny:2022ud}.

In ultraclean electronic channels, a significant modification of the anomalous Hall effect is expected~\cite{PhysRevLett.121.226601,PhysRevResearch.3.033075,PhysRevB.104.184414,PhysRevB.103.125106,Glazov_2021b,PhysRevB.106.L041407}. In recent works~\cite{Glazov_2021b,grigoryan2023anomalous}, within the framework of the kinetic equation for the spin density matrix, a theory of the spin and anomalous Hall effects in ultraclean channels with a two-dimensional electron gas has been developed and the contributions of microscopic mechanisms have been studied in case where momentum relaxation occurs on the edges of the channel, and the field acts on electrons over the entire area of the structure. At the same time, the cases of ballistic transport, when $l_{ee} \gg w$ (and interparticle collisions practically do not play a role), and hydrodynamic transport, when $l_{ee} \ll w \ll l$, were considered separately. The purpose of this article is to develop the theory of transition between hydrodynamic and diffusion regimes of electron transport in the anomalous Hall effect.
Here we  consider an electron gas with a parabolic spectrum and assume an arbitrary relation between $l$ and $l_{ee}$, but assume that $l_{ee} \ll w$, i.e. that electron-electron collisions are very efficient.

In Sec.~\ref{sec:model} we present a model of the anomalous Hall effect within the framework of the spin density matrix method and present the general form of the kinetic equation. Section~\ref{sec:norm} discusses the normal Hall effect at the transition from the diffusive to the hydrodynamic one. Section~\ref{sec:AHE} is devoted to the anomalous Hall effect, it contains the  main results of the work. The results are summarized in section~ \ref{sec:conclusion}.

\section{\label{sec:model}Model}

\begin{figure}[b]
\includegraphics[width=0.45\linewidth]{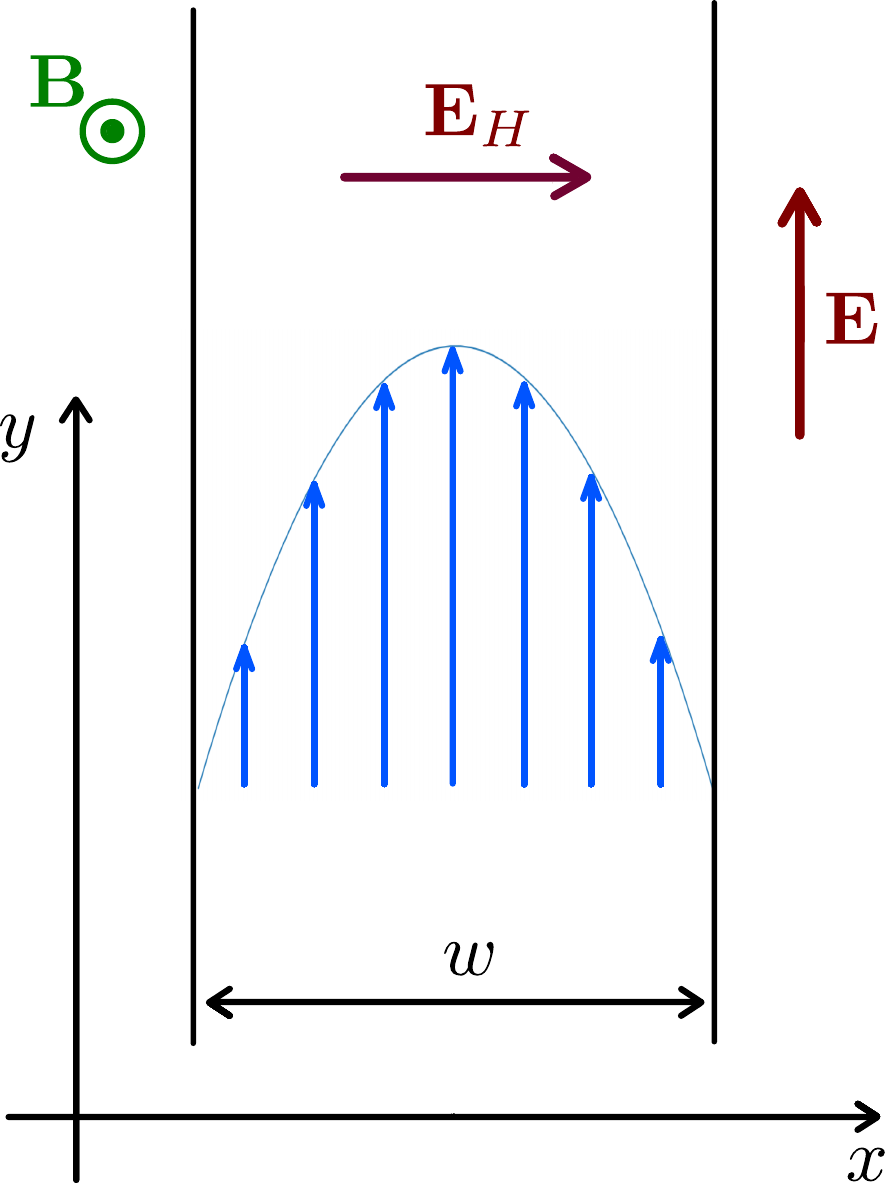}
\caption{Schematic representation of the system under study. The external electric field $\mathbf E$ is directed along the channel axis $y$, the external magnetic field $\mathbf B$ is directed along the normal to the structure $z$. As a consequence, a Hall field $\mathbf E_H$ appears, directed along $x$. The arrows and parabolic curve show the electric current profile in the channel in case of hydrodynamic regime.}
\label{fig:geom}
\end{figure}

The geometry of the system is presented in Fig.~\ref{fig:geom}. An external electric field is applied along the axis of the channel with a quasi-two-dimensional electron gas, $\mathbf E\parallel y$ (in this direction the channel is assumed to be infinite, electric current can flow along $y$), the channel width along $x$ is equal to $w$, the external magnetic field is applied perpendicular to the plane of the two-dimensional electron gas $\mathbf B\parallel z$. Scattering on the channel edges is assumed to be diffusive.

We describe electron transport in the channel in terms of the kinetic equation for the spin density matrix
\begin{equation}
     \hat \rho_{\mathbf p} = f_{\mathbf p} \hat I + \mathbf s_{\mathbf p} \cdot \hat{\bm \sigma},
\end{equation}
where $\mathbf p$ is the electron momentum, $\hat I$ is the identity matrix of size $2\times 2$, $\hat{\bm \sigma}=(\hat \sigma_x,\hat \sigma_y,\hat\sigma_z)$ is a vector composed of Pauli matrices, $f_{\mathbf p}=\Tr\{\hat\rho_{\mathbf p}/2\}$ is a spin-averaged electron distribution function, $\mathbf s_{\mathbf p} = \Tr\{\hat\rho_{\mathbf p} \hat{\bm \sigma}/2 \}$ is a spin distribution function. The density matrix satisfies the kinetic equation~\cite{grigoryan2023anomalous}
\begin{multline}\label{eq:kin:0}
   \frac{\partial }{\partial \bold{r}}\hat{\mathbf v}_{\mathbf p} \hat\rho_\bold{p}  +e(\bold{E}+ \mathbf E_H)\frac{\partial \hat\rho_{\bold{p}}}{\partial \bold{p}} + \frac{e}{c}[\bold{v}\times\bold{B}]\frac{\partial \hat\rho_{\bold{p}}}{\partial \bold{p}} 
   \\
   =- \frac{\hat\rho_{\bold{p}}-\overline{\hat\rho_\bold{p}}}{\tau} + \hat{Q}_{ee}\{\hat\rho_{\mathbf p}\} + \hat{G}_{\mathbf p},
\end{multline}
where $\hat{\mathbf v}_{\mathbf p}$ is the electron velocity operator (taking into account the anomalous contributions given in Sec.), $\mathbf E_H$ is the Hall field arising due to the redistribution of electrons induced by the Hall effect , $e < 0$ is the electron charge, $c$ is the speed of light, $\overline{\hat \rho_{\mathbf p}} = (2\pi)^{-1}\int_0^{2\pi}\hat \rho_{\mathbf p} d\varphi$ is the density matrix averaged over the angles of the vector $\mathbf p$, $\tau$ is the electron scattering time due to electron-impurity and electron-phonon collisions (momentum relaxation time), $Q_{ee}\{\hat\rho_{\mathbf p}\}$ is the electron-electron collision integral, $\hat G_{\mathbf p}$ is the generation rate of anomalous Hall current. Here and below, $\partial /\partial \mathbf{r}$ and $\partial/\partial\mathbf{p}$ denote gradients in spatial coordinate and momentum. We neglect Fermi-liquid renormalizations.

The kinetic equation~\eqref{eq:kin:0} should be supplemented with the boundary conditions. For the diffusive scattering on the channel edges, these conditions have the form
\begin{equation}\label{eq:boundary conditions}
     \hat \rho_{\mathbf{p}}(\pm w/2) = \begin{cases}
     \text{const},\; p_x>0,\; x=-w/2,\\
     \text{const},\; p_x<0,\; x=w/2,
     \end{cases}
\end{equation}
where $p_x = p\cos\varphi$ is the $x$-component of the momentum. The physical sense of the boundary condition~\eqref{eq:boundary conditions} is that electrons after scattering on the edges have an isotropic distribution function. Also, we require that the electron flux through the edges is zero, $\sum_{\mathbf p}\hat{\mathbf{v}}_{\mathbf p,x}\hat{\rho}_{\mathbf p}=0$.

We are interested in case of degenerate electron gas, where $T \lesssim \varepsilon_F$, where $\varepsilon_F$ is the Fermi energy, and the Boltzmann constant is set equal to unity. Let us introduce the energy-integrated particle and spin distribution functions, which depend only on the angle between $\mathbf p$ and the $x$ axis, namely,
\begin{equation}
     F_\varphi = \mathcal D \int_0^\infty f_{\mathbf p}d\varepsilon, \;\;\; \mathbf S_\varphi = \mathcal D \int_0^\infty \mathbf s_{\mathbf p}d\varepsilon,
\end{equation}
where $\mathcal D = m/2\pi\hbar^2$ is the density of states per spin, $\varepsilon = p^2/2m$ is the electron spectrum, which we consider to be parabolic, $m$ is the electron effective mass.

Since the Hall field is parallel to the $x$ axis (see Fig.~\ref{fig:geom}), we introduce the electrostatic potential $\Phi(x)$ according to
\begin{equation}\label{Hall:E}
     \mathbf{E}_H = -\frac{\partial \Phi(x)}{\partial x}\hat{\mathbf x},
\end{equation}
where $\hat{\mathbf x}$ is the unit vector in the $x$ direction. This potential can be included in the renormalized distribution function by defining $\tilde{F}_\varphi(x)$ as
\begin{equation}\label{ren:F}
     \tilde{F}_\varphi(x) = F_\varphi(x) + e\mathcal D \Phi(x).
\end{equation}
Under typical conditions, the electric current caused by the imbalance of charges significantly exceeds the diffusive current caused by the gradient of the chemical potential, therefore~\cite{PhysRevB.98.165440,grigoryan2023anomalous}
\begin{equation}\label{Hall:E approx}
     \mathbf E_H \approx -\frac{1}{e\mathcal D}\frac{\overline{\partial\tilde F_\varphi}}{\partial x} \hat{\mathbf x}.
\end{equation}
Our goal is to find the Hall field in the linear-$\mathbf E$ and linear-$\mathbf B$ approximation. Taking the trace of the equation \eqref{eq:kin:0} and integrating it over energy taking into account the replacement \eqref{ren:F}, we obtain the following kinetic equation:
\begin{multline}\label{eq:kin1}
    \frac{\partial}{\partial x} \left(v_x \tilde F_\varphi + v_aS_{z,\varphi}^0\right) + \omega_c\frac{\partial \tilde  F_\varphi}{\partial \varphi} + \frac{\tilde F_\varphi - \overline{\tilde F_\varphi}}{\tau}  \\
     =Q_{ee}\{\tilde F_\varphi\} + e\mathcal D Ev_y + G_{\varphi}.
\end{multline}
Here $v_x = v\cos\varphi, \;\;\; v_y = v\sin\varphi$ are the velocity components, where $v = \sqrt{2\varepsilon_F/m}$ is the Fermi velocity, $\omega_c = -eB/mc$ is the cyclotron frequency, $v_a \propto E$ is the anomalous part of the velocity (see section~\ref{sec:AHE}), and 
\begin{equation}
\label{equilibriumSz}
    S_{z,\varphi}^0 \equiv S_z^0 = -\frac{1}{2} \mathcal D g \mu_BB
\end{equation}
is equilibrium spin polarization in a magnetic field, $g$ is the $g$-factor of the electron, $\mu_B$ is the Bohr magneton.

\section{Normal Hall effect}\label{sec:norm}
In this section, to illustrate the general approach, we briefly discuss the kinetic equations for the ordinary and spin distribution functions and their solutions without taking into account anomalous contributions. Here we describe the normal Hall effect caused by the action of the Lorentz force on electrons. We also analyze how the Hall field changes during the transition from the diffusive to the hydrodynamic regimes of transport.

\subsection{Distribution function and Hall field}
The kinetic equation \eqref{eq:kin1}, neglecting anomalous contributions, takes the form
\begin{equation}\label{eq:kin normal}
    \frac{\partial}{\partial x} \left(v_x \tilde F_\varphi\right) + \omega_c\frac{\partial \tilde  F_\varphi}{\partial \varphi} + \frac{\tilde F_\varphi - \overline{\tilde F_\varphi}}{\tau}  
    =Q_{ee}\{\tilde F_\varphi\} + e\mathcal D Ev_y .
\end{equation}
To describe electron-electron collisions, we use the relaxation time approximation. Then the integral of interparticle collisions is written as~\cite{PhysRevB.100.125419,PhysRevB.98.165412,Glazov_2021b,PhysRevB.104.085434,Narozhny:2022ud,grigoryan2023anomalous}
\begin{equation}
    \label{Qee}
    Q_{ee}\{ \tilde F_\varphi\} = \frac{\tilde F_\varphi - \overline{\tilde F_{\varphi}}- \overline{\tilde F_{\varphi}^c}\cos{\varphi}- \overline{\tilde F_{\varphi}^s}\sin{\varphi}}{\tau_{ee}},
\end{equation}
where
\begin{equation}
    \label{Fcs}
    \overline{\tilde F_{\varphi}^c} = \frac{1}{\pi}\int_0^{2\pi} \tilde F_{\varphi}\cos{\varphi}\; d\varphi, \quad    \overline{\tilde F_{\varphi}^s} = \frac{1}{\pi}\int_0^{2\pi} \tilde F_{\varphi}\sin{\varphi} \; d\varphi.
\end{equation}
Here $\tau_{ee}$ is the effective time of electron-electron collisions. We emphasize that this simplified form of the electron-electron collision integral takes into account the conservation of the number of particles (the zero harmonic of the distribution function harmonic) and momentum (the first harmonic of the distribution function).

\begin{figure}[b]
\includegraphics[width=0.99\linewidth]{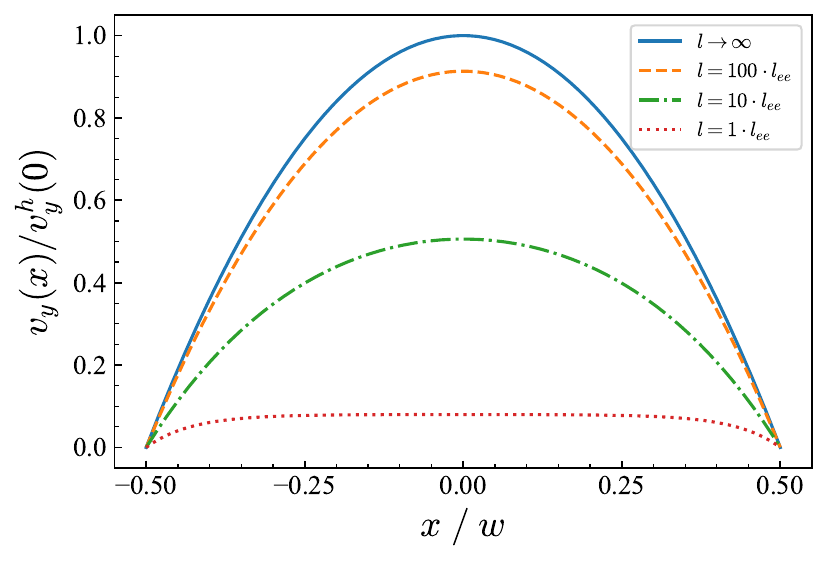}
\caption{
Coordinate dependence of electron velocity $v_y \propto \delta F_1(x)$ in \eqref{F harmonics E}, calculated using the Eqs.~\eqref{eq: solutions on harmonics} and \eqref{lambdas} for various ratios $l/l_{ee}$ indicated in the legend. Velocity normalized to its value in the  hydrodynamic regime (at $l\to\infty$) $v_y^h(0) = eE\mathcal D w^2/4\tau_{ee}$.}
\label{fig:v normal}
\end{figure}

\begin{figure*}[t]
    \centering
    \includegraphics[width=0.98\textwidth]{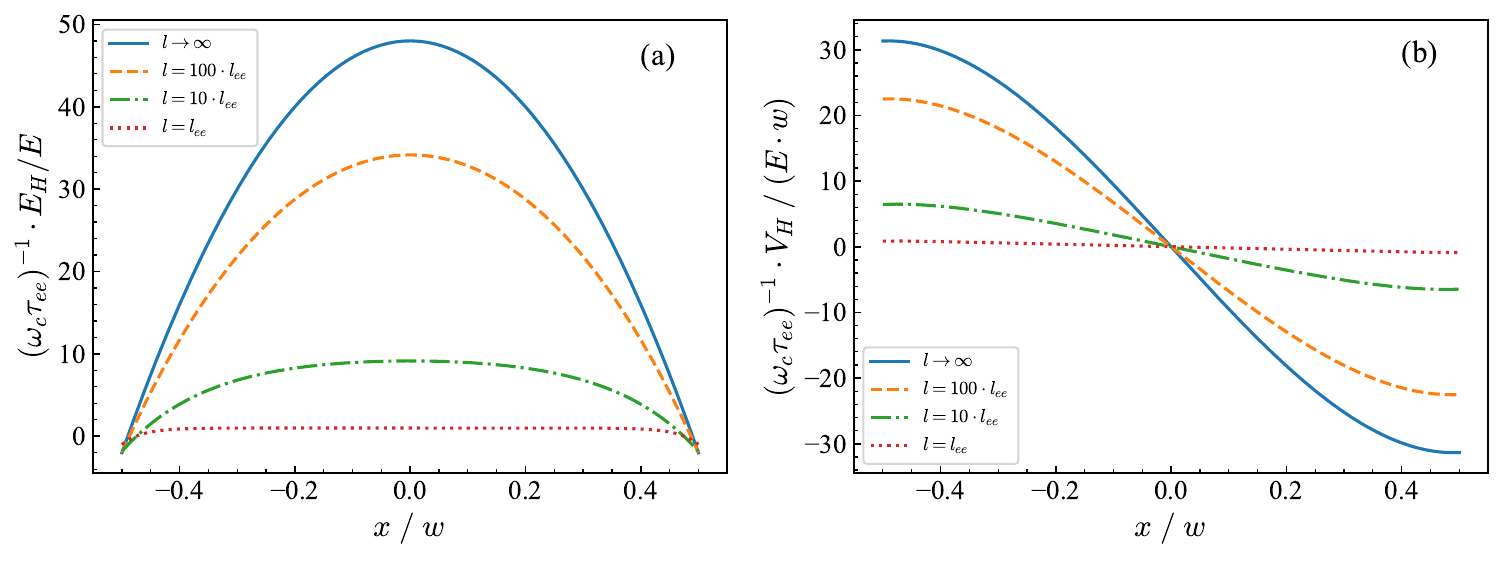}
    \caption{Coordinate dependence of the Hall field (a) and voltage (b) during the transition from the diffusive to hydrodynamic regime in the normal Hall effect.}
    \label{fig:Hall normal}
\end{figure*}

In our case where $l_{ee}/w \ll 1$, $l/l_{ee}$ is arbitrary, we can solve the kinetic equation by expanding the integral distribution function and taking into account only angular harmonics up to the second order, at any ratio $l/l_{ee}$. Assuming the magnetic field to be weak $\omega_c \tau_{ee} \ll 1$, we will look for a solution to the equation~\eqref{eq:kin normal} by iterations over the magnetic field. Setting $\mathbf B=0$, we write the distribution function in the form\begin{equation}
    \label{F harmonics E}
    \delta \tilde F_{\varphi}(x) = \delta F_1(x) \sin{\varphi} + \delta F_2(x) \sin{2\varphi} \propto E,
\end{equation}
Substituting the expansion \eqref{F harmonics E} and the collision integral \eqref{Qee} into the kinetic equation \eqref{eq:kin normal}, we obtain the equations on angular harmonics
\begin{subequations}
     \label{eq: eq.s on harmonics}
\begin{align}
         \frac{l_1}{2}\frac{\partial \delta F_{2}}{\partial x}+\delta F_{1}=eEl_1 \mathcal D,\\
         \frac{l_{2}}{2}\frac{\partial \delta F_{1}}{\partial x}+\delta F_{2}=0,
\end{align}
\end{subequations}
where $l_{1,2} = v\tau_{1,2}$, and $\tau_{1} \equiv \tau$, $\tau_2 = (\tau^{-1} +\tau_{ee} ^{-1})^{-1}$, respectively, are the relaxation times of the first and second angular harmonics. Solving this system taking into account the condition of vanishing current at the boundary,\footnote{For $l_{ee} \ll w$ the condition~\eqref{eq:boundary conditions} is reduced to the requirement that the electron velocity at the boundary vanishes.} $\delta F_1 (x = \pm w/2) = 0$, we get
\begin{equation}
       \label{eq: solutions on harmonics}
       \delta F_{1,2} = eE \mathcal D l_1 \lambda_{1,2}(x),
\end{equation}
with functions
\begin{subequations}
     \label{lambdas}
\begin{align}
        \lambda_1(x)= l_1\left(1 - \frac{\cosh\left(2x/\sqrt{l_1l_2}\right)}{\cosh\left(w/\sqrt{l_1l_2}\right)}\right ),\label{lambdas1a}
          \\
          \lambda_2(x) = \sqrt{l_1l_2}\frac{\sinh\left(2x/\sqrt{l_1l_2}\right)}{\cosh\left(w/\sqrt{l_1l_2}\right)}.
\end{align}
\end{subequations}
In the limit $l\rightarrow \infty$ or, more precisely, at $ll_{ee} \gg w^2$, the hydrodynamic regime of electron transport and the Poiseuille flow are realized, characterized by a parabolic dependence of the electron velocity on the coordinate (Fig.~\ref{fig:geom}). In the limit $l \ll w$ the diffusive regime is realized, and the flow profile becomes flat and does not depend on the coordinate (strictly speaking, the velocity vanishes near the walls on a negligibly small scale $\sim l_{2}$). The transition between these regimes is illustrated in Fig.~\ref{fig:v normal}, which shows the velocity profile calculated according to~\eqref{eq: solutions on harmonics} and \eqref{lambdas} for different ratios $l/l_{ee} $. It can be seen that as the mean free path $l$ decreases, the electron velocity decreases, and the spatial distribution of velocities changes qualitatively: from parabolic at $l\to \infty$, which corresponds to the Poiseuille flow, to flat at $l\to 0$, which corresponds to diffusive regime.

Now, let us take into account the magnetic field and find the Hall field. We substitute the solution in presence of only the electric field \eqref{eq: solutions on harmonics} into the Lorentz term with $\omega_c$ in \eqref{eq:kin normal} and determine the linear-$\mathbf B$ contribution to the integral distribution function. The expansion of this magnetic field induced contribution into angular harmonics will be as follows
\begin{equation}
     \label{F harmonics B}
     \Delta \tilde F_{\varphi}(x) = \Delta \tilde F_0(x) + \Delta F_1(x) \cos{ \varphi} + \Delta F_2(x) \cos{2\varphi},
\end{equation}
and the coefficients for angular harmonics satisfy the equations
\begin{subequations}
     \label{eq: eq on harmonics B normal}
\begin{align}
         \frac{v}{2}\frac{\partial \Delta F_1}{\partial x} = 0, \\
         v\frac{\partial \Delta\tilde F_0}{\partial x} + \frac{v}{2}\frac{\partial \Delta F_2}{\partial x} + \omega_c\delta F_1 + \frac{ \Delta F_1}{\tau} = 0, \\
         \frac{v}{2}\frac{\partial \Delta F_1}{\partial x} + 2\omega_c\delta F_2 + \frac{\Delta F_2}{\tau_2} = 0.
\end{align}
\end{subequations}
Solving this set of equations and using the relation \eqref{Hall:E approx}, we obtain an expression for the normal Hall field:
\begin{equation}
\label{eq:Hall:norm:EH}
     E_H = E\cdot \omega_c\tau_2\left(\frac{\lambda_1}{\l_2} - \frac{\partial \lambda_2}{\partial x}\right),
\end{equation}
where the functions $\lambda_1$, $\lambda_2$ are introduced in \eqref{lambdas}. Figure~\ref{fig:Hall normal}(a) shows Hall field $E_H(x)$, calculated using the equation~\eqref{eq:Hall:norm:EH} for various ratios $l/l_{ ee}$. Panel (b) of Fig.~\ref{fig:Hall normal} shows the Hall voltage obtained by integrating $E_H$ into~\eqref{eq:Hall:norm:EH}, for the same values of $l/l_{ee}$. In the limiting case of the diffusive regime $E_H = \omega_c \tau_1 E$, and in the hydrodynamic regime, when $l l_{ee} \gg w^2$ and the Hall field has a nontrivial coordinate dependence in accordance with Ref.~\cite{PhysRevB.106.245415}.

\subsection{Spin distribution function}
Now let us determine the spin distribution function $s_{z,\mathbf p}$, which is necessary for further calculation of the anomalous Hall effect (cf. Ref.~\cite{grigoryan2023anomalous}). As above, to calculate the spin distribution function we neglect the anomalous contributions to the kinetic equation and take into account only the relaxation of the spin to the equilibrium one, Eq.~\eqref{equilibriumSz}.
Let us write down the kinetic equation for the spin distribution function integrated over energy, expanding it similarly to the ordinary distribution function, up to the second angular harmonic:
\begin{equation}\label{eq:kin:sz}
    \cos\varphi \frac{\partial S_{z,\varphi}}{\partial x} + \frac{S_{z}^1\sin\varphi}{\tau_1^s} + \frac{S_{z}^2\sin2\varphi}{\tau_2^s} = e\mathcal D E\frac{S_z^0}{N}\sin\varphi,
\end{equation}
where $S_z^1, \; S_z^2$ are expansion coefficients of the first and second angular harmonics respectively, $\tau_1^s, \; \tau_2^s$ are their relaxation times, $N$ is the average electron density of the system.

It is important to note that the relaxation times of the spin distribution function depends on the magnetic field. In the case of a low magnetic field, where
\begin{subequations}
    \label{B:limits}
\begin{equation}\label{lowB}
    |g\mu_BB|\ll T\ll \varepsilon_F,
\end{equation}
relaxation of both the first and second harmonics occurs due to both electron-impurity and electron-electron collisions, with $\tau_1^s = \tau_2^s = (\tau^{-1}+\tau_{ ee}^{-1})^{-1}$~\cite{glazov02,amico:045307,wu03prb,glazov04a}. As we will see later, in this situation, the diffusive regime is always realized, as $l_{ee} \ll w$. For moderate magnetic fields, when
\begin{equation}\label{modB}
    T \ll  |g\mu_BB| \ll \varepsilon_F,
\end{equation}
\end{subequations}
relaxation of the first harmonic during electron-electron collisions is suppressed~\cite{alekseev:2}, as collisions between electrons with opposite spins become extremely ineffective. In this case, $\tau_1^{s} = \tau$ and $\tau_2^{s} = (\tau^{-1}+\tau_{ee}^{-1})^{-1}$.

Solution of Eq.~\eqref{eq:kin:sz} is similar to determination of $\delta F_{1,2}$ from Eqs.~\eqref{eq: eq.s on harmonics}. By analogy with Eq.~\eqref{eq: solutions on harmonics} we obtain
\begin{equation}
        \label{eq: solutions on spin harmonics}
        S_{z}^{1,2}(x) = \frac{S_z^0}{N}eE\mathcal D \lambda_{1,2}^s(x),
\end{equation}
where the functions $\lambda_{1,2}^2$ differ from the functions $\lambda_{1,2}$ in ~\eqref{lambdas} by replacing the relaxation lengths $l_{1,2}$ with ``spin'' ones $ l_{1,2}^s = v \tau_{1,2}^s$.

In the case of a low magnetic field \eqref{lowB} the second harmonic \eqref{eq: solutions on spin harmonics} is zero everywhere except for a narrow regions with width $l_2$ near the channel edges. These stripes do not make a noticeable contribution to the anomalous Hall effect, so we will neglect it in follows. The first angular harmonic \eqref{eq: solutions on spin harmonics} in this limit does not depend on the coordinate. This result corresponds to the diffusion mode of spin transport, despite the fact that $l\gg l_{ee}$.

In moderate magnetic fields \eqref{modB} both harmonics remain and, generally speaking, significantly depend on the coordinate. In this case, the evolution of the $S_z^1$ profile with a change in the ratio $\tau/\tau_{ee}$ is similar to that shown in Fig.~\ref{fig:v normal} for the spatial distribution of velocities.

\section{Anomalous Hall effect}\label{sec:AHE}
\subsection{Model}

It is known that anomalous Hall effect is caused by spin-orbit interaction. There are three main mechanisms of the effect~\cite{Sinitsyn_2007}: anomalous velocity, shifts of electron wave packets caused by scattering on impurities, and asymmetric electron scattering. These three mechanisms are accounted for in what follows. Following the works~\cite{2020arXiv200405091G,Glazov_2021b,grigoryan2023anomalous}, where these mechanisms are discussed in detail, we present expressions for the corresponding contributions to the kinetic Eq.~\eqref{eq:kin1}.

Let us start with mechanisms whose contributions does not depend on the transport regime and coordinate. The first one is \emph{anomalous velocity}, that arises in an electric field for spin-polarized electrons in the presence of non-zero Berry curvature of energy bands. It has a different sign for electrons with different spins and included in the velocity operator in the kinetic equation \eqref{eq:kin:0} as:
\begin{equation}\label{eq:v_a operator}
    \hat{\mathbf v}_{a,B} = \hat{\sigma}_z \mathbf v_{a,B}, \quad \mathbf v_{a,B} = -\frac{2\xi e} {\hbar}[\bold{\hat{z}\times E}],
\end{equation}
where $\hat{z}$ is the unit-vector perpendicular to the structure, and the subscript $B$ in $V_{a,B}$ means that this is a contribution to the anomalous velocity due to the presence of Berry curvature. The parameter $\xi$ characterizes the strength of the spin-orbit coupling~\cite{2020arXiv200405091G,Glazov_2021b,grigoryan2023anomalous}. The corresponding contribution to the anomalous velocity in the equation~\eqref{eq:kin1} can be represented as
\begin{equation}\label{eq:v_a}
     v_{a,B} = \frac{2\xi e}{\hbar}E.
\end{equation}

The second coordinate and transport regime independent contribution is due to the anomalous distribution of electrons that arises due to scattering on impurities or phonons, taking into account the scattering asymmetry induced by shifts of wave packets: the contribution of \emph{anomalous distribution}. The anomalous distribution effect contributes to the generation term in the kinetic Eq.~\eqref{eq:kin1}
\begin{equation}\label{eq:Gadist}
     G_{\varphi,adist} = G_{adist}\cos\varphi,
\end{equation}
where
\begin{equation}
     \label{eq:Gadist:1}
      G_{adist}=-\left(1+\nu\right)\frac{2\xi e}{l\hbar}E S_z^{0},
\end{equation}
and $\nu$ is a parameter that depends on the scattering mechanism~\cite{2020arXiv200405091G,grigoryan2023anomalous}.
Shifts of wave packets during scattering also lead to a contribution in anomalous velocity, called the side jump accumulation. The corresponding anomalous contribution to the velocity in \eqref{eq:kin1} depends on the coordinate
\begin{equation}\label{eq:v_sj}
  v_{a,sj}(x) = -\left(1+\nu\right)\frac{\lambda_1^s(x)}{l}\frac{\xi}{\hbar} eE,
\end{equation}
where the function $\lambda_1^s(x)$ describes the spatial profile of the first angular harmonic of the spin distribution function, see Eqs.~\eqref{eq: solutions on spin harmonics} and \eqref{lambdas1a}.

The last mechanism of the anomalous Hall effect is associated with \emph{asymmetric scattering (skew scattering)} on impurities. It makes the following contribution to the generation term in Eq.~\eqref{eq:kin1}
\begin{equation}\label{eq:Gsk}
     G_{\varphi,sk}(x) = G_{sk}\cos\varphi,
\end{equation}
where
\begin{equation}\label{eq:Gsk:1}
     G_{sk} = S_{imp}\frac{\tau^2}{\hbar}\frac{\lambda_1^s(x)}{l}\frac{\xi e}{l\hbar}E\frac {N S_z^{0}}{2},
\end{equation}
and the coefficient $S_{imp}$ determines the degree of scattering asymmetry. The expression for $S_{imp}$ according to~\cite{2020arXiv200405091G} (see also~\cite{Ado_2015}) includes two contributions~\footnote{The expression~\eqref{eq:Simp} corrects a typo in the formula (26) from~\cite{Glazov_2021b}.}
\begin{equation}\label{eq:Simp}
     S_{imp} = \frac{2\pi U_v}{\tau} + \frac{8\nu \hbar}{N\tau^2},
\end{equation}
where in the case of scattering by short-range defects $U_v$ is the Fourier transform (power) of the potential of a single defect. The first term in \eqref{eq:Simp} corresponds to the third-order contribution in the defect potential, and second term corresponds to the contribution to two-impurity coherent scattering~\cite{2020arXiv200405091G,grigoryan2023anomalous}. We emphasize that electron-electron collisions conserve the total momentum of the pair, so they cannot  contribute to the anomalous Hall effect, but can lead to the generation of spin currents.

Since in the linear-$\mathbf B$ approximation the normal and anomalous contributions to the Hall effect are additive, the anomalous part of the distribution function (responsible for the anomalous Hall effect) satisfies the following equation
\begin{multline}\label{eq:kin anomalous}
    \frac{\partial}{\partial x} \left[v_x \tilde F_\varphi + (v_{a,B} + v_{a,sj})S_{z,\varphi}^0\right] + \frac{\tilde F_\varphi - \overline{\tilde F_\varphi}}{\tau}  \\ 
    =Q_{ee}\{\tilde F_\varphi\} + (G_{adist}+G_{sk})\cos{\varphi}.
\end{multline}
This equation is obtained from~\eqref{eq:kin1} by eliminating the contributions containing the external electric field $e\mathcal D E v_y$ and the Lorentz force $\omega_c \partial \tilde F_\varphi/\partial\varphi$. As in Sec.~\ref{sec:norm}, the distribution function, can be expanded into three angular harmonics:
\begin{equation}\label{expension hydro}
     \tilde{F}_\varphi = \overline{\tilde{F}_\varphi}+F_1\cos\varphi + F_2\cos2\varphi.
\end{equation}
The coefficients for angular harmonics satisfy the system of equations [cf. Eq.~\eqref{eq: eq on harmonics B normal}]
\begin{subequations}
     \begin{equation}\label{eq. system anomalous hydro 0}
         \frac{\partial}{\partial x}\left(\frac{v}{2} F_1 + v_a S_z^0\right) = 0,
     \end{equation}
     \begin{equation}\label{eq. system anomalous hydro 1}
         \frac{\partial}{\partial x}\left[v\left(\overline{\tilde{F}_\varphi}+\frac{F_2}{2}\right)\right] = -\frac{ F_1}{\tau_1} + G,
     \end{equation}
     \begin{equation}\label{eq. system anomalous hydro 2}
         \frac{\partial}{\partial x}\left(\frac{v}{2}F_1\right) + \frac{F_2}{\tau_2}=0,
     \end{equation}
\end{subequations}
Generalizing the solution obtained in \cite{grigoryan2023anomalous} in the hydrodynamic regime, and taking into account the relation \eqref{Hall:E approx}, we obtain the expression for the anomalous Hall field:
\begin{equation}\label{AHE:E_H}
     E_{H,a} = -\frac{2S_z^0}{ev\mathcal D}\frac{v_a}{l_1}+\frac{l_{2}}{2ev\mathcal{D}}\frac{\partial^2}{\partial x^2}(v_a S_z^0)-\frac{G}{ev\mathcal D}.
\end{equation}
where $v_a = v_{a,B} + v_{a,sj}$, $G = G_{adist} +G_{sk}$. Equation~\eqref{AHE:E_H} is the key result of our work; it generalizes the formulas obtained in~\cite{grigoryan2023anomalous} to the case of an arbitrary relation between $l$ and $l_{ee}$.

\subsection{Analysis of results}

Now we turn to the analysis of the obtained results. We will consider two cases of low and moderate magnetic fields, where the condition~\eqref{lowB} or \eqref{modB} is satisfied, respectively.
\begin{figure*}[t]
     \centering
     \includegraphics[width=0.98\textwidth]{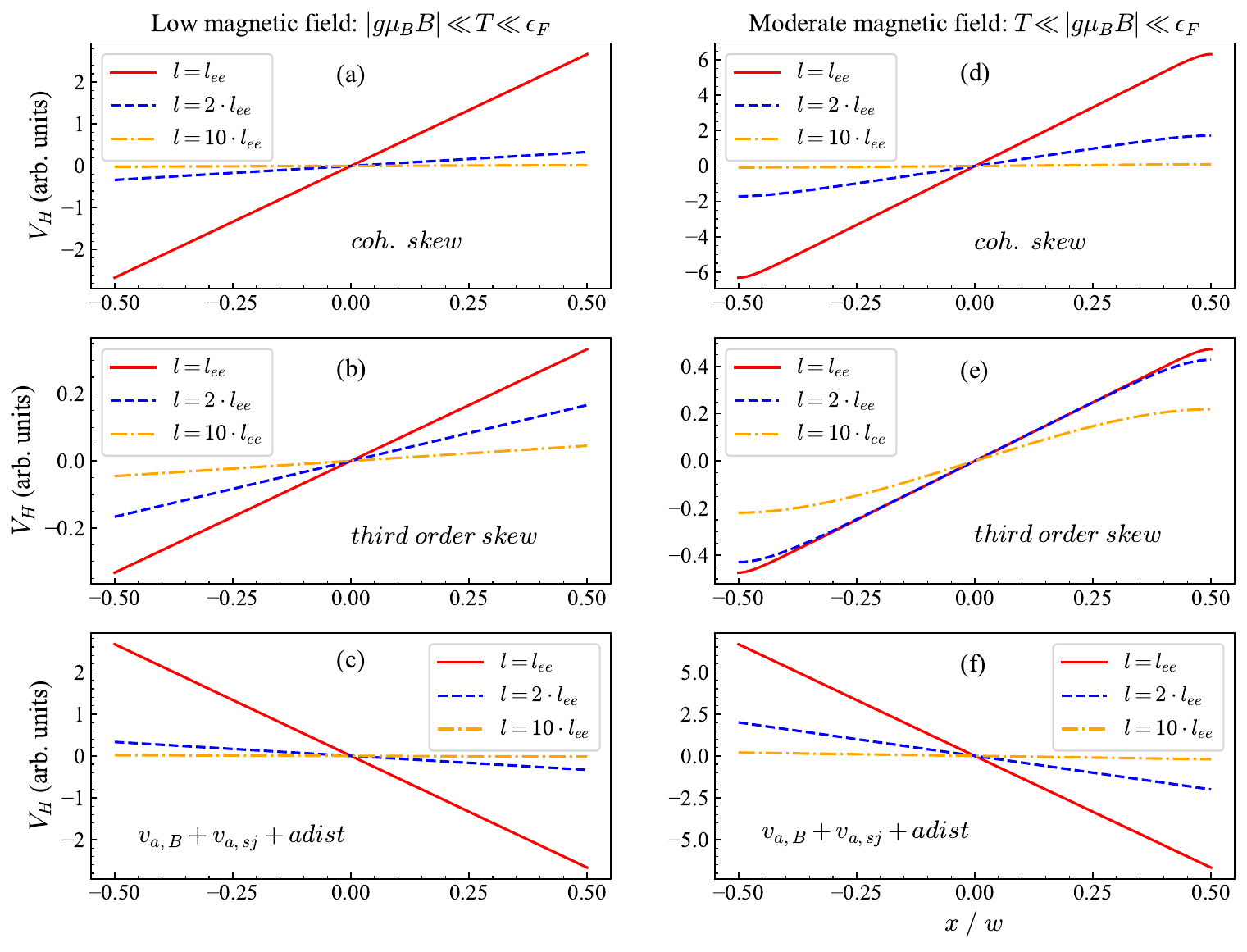}
     \caption{Coordinate dependencies of the contributions of coherent skew scattering (a,d), third-order skew scattering (b,e) and anomalous contributions (c,f) to the anomalous Hall voltage for different $l/l_{ee}$ ratios in case of low [panels (a-c)] and moderate [panels (b-f)] magnetic fields, calculated using the expression~\eqref{AHE:E_H}. In this calculation, $\nu = 1$, $\pi U_v N / \tau_{ee}\hbar = 1, \; w/l_{ee} = 5$.}
     \label{fig:Hall anomalous}
\end{figure*}

\subsubsection{Low magnetic field}
In case of low magnetic fields, where Eq.~\eqref{lowB} is satisfied, the first harmonic of the spin distribution function effectively relaxes due to interparticle collisions. For the spin distribution function in this case, the diffusive regime is realized regardless of the relationship between $l$ and $l_{ee}$, and the contributions of all mechanisms of the anomalous Hall effect turn out to be coordinate independent. The quantities $v_{a,sj}$ and $G_{\varphi,sk}$ take the form (44a) and (50a) from \cite{grigoryan2023anomalous}. For $l \rightarrow \infty$ from \eqref{AHE:E_H} we obtain for the anomalous contributions formulas (68a), (68b) and (68d) from \cite{grigoryan2023anomalous}. For an arbitrary $l/l_{ee}$, the sum of all contributions will be as follows:
\begin{equation}\label{AHE: lowB}
    E_{H,a} = -\frac{S_z^0E}{ev\mathcal D}\cdot \frac{2\xi e}{\hbar (l+l_{ee})}\left(1 - \nu + \frac{\pi U_vN}{2v
    \hbar}l_{ee}\right).
\end{equation}

The anomalous Hall voltage
\[
V_{H,a} = - \int_0^x E_{H,a}(x')dx',
\]
[cf. Eq.~\eqref{Hall:E}] turns out to be a linear function of the coordinate in this case. The coordinate dependencies of individual contributions to the anomalous Hall voltage are shown in panels (a) and (b) of Fig.~\ref{fig:Hall anomalous}, respectively, for two-impurity coherent scattering and scattering on impurities in the third order in the impurity potential. The sum of contributions from other mechanisms to the Hall voltage is shown in Fig.~\ref{fig:Hall anomalous}(c).
Depending on the main mechanism of the anomalous Hall effect, the signs of the parameters $\xi$, $\nu$ and $U_v$, the anomalous Hall field, Eq.~\eqref{AHE: lowB}, can be either collinear or opposite to the normal one, Eq.~\eqref{eq:Hall:norm:EH}, and does not change direction when passing from the limiting case $l\rightarrow 0$ to the limiting case $l\rightarrow \infty$ (for a fixed $l_{ee}$).

\begin{figure}[tb]
\includegraphics[width=0.99\linewidth]{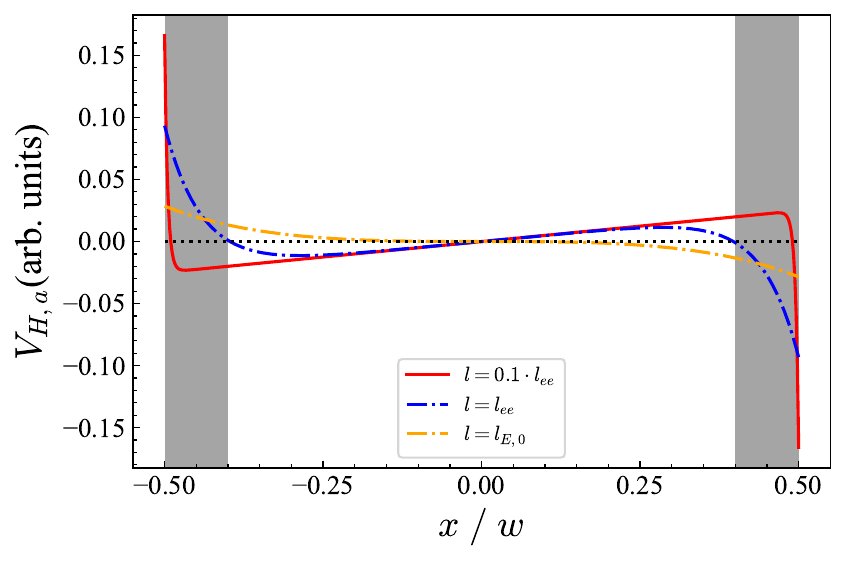}
\caption{Coordinate dependence of anomalous Hall voltage found from the total Hall field [sum of Eqs.~\eqref{AHE: modB v_a+v_sj+adist} and \eqref{AHE: modB sk}], in the case of moderate magnetic fields. The gray stripes indicate the regions of width $\sim l_{ee}$ near the edges where the three harmonics approximation is invalid. At $l = 0.1\cdot l_{ee}$ (red curve), the voltage increases permanently and linearly in the sample, which corresponds to the deeply diffuse regime, Eq.~\eqref{AHE:mod:B:diff}. For $l = l_{ee}$, the Hall voltage is a non-monotonic function of the coordinate. At the value $l = l_{E,0}$, the Hall field at the point $x=0$ vanishes and the Hall voltage again becomes a monotonic function of the coordinate. For $l \gtrsim l_{E,0}$ remains a monotonic function with a third order coordinate dependence. In the calculation $\pi U_v N \tau_{ee}/\nu\hbar = 0.05,\; w/l_{ee} = 10$. In case of such parameters $l_{E,0} / l_{ee} \approx 19.941$.
\label{fig:AHE sign transition}
}
\end{figure}
\subsubsection{Moderate magnetic fields}
In the case of moderate magnetic fields, where the condition \eqref{modB} is satisfied, collisions of electrons with opposite spins are suppressed. In this case, the contributions from $v_{a,B}$ and $G_{adist}$ are the same as in the previous case, since they do not depend on the transport regime. The anomalous velocity associated side jump accumulation $v_{a,sj}$ in coordinate dependent, however, the sum of the contributions from the first and second terms in Eq.~\eqref{AHE:E_H} turns out to be coordinate independent. Finally, for the sum of contributions from anomalous velocities and anomalous distribution in case of moderate magnetic field, we have
\begin{equation}\label{AHE: modB v_a+v_sj+adist}
     \tilde{E}_{H,a,B} + \tilde{E}_{H,a,sj} + \tilde{E}_{H,adist} = \frac{4E S_z^0 }{ev \mathcal D}\frac{\xi eE}{\hbar l}\nu,
\end{equation}
The Hall voltage dependence corresponding to the field in \eqref{AHE: modB v_a+v_sj+adist} is shown in panel (f) of Fig.~\ref{fig:Hall anomalous}.
Note that the contributions of asymmetric scattering associated with both two-impurity coherent scattering and third-order asymmetric scattering on single impurities have a nontrivial coordinate dependence:
\begin{equation}\label{AHE: modB sk}
    \tilde{E}_{H,sk,coh} + \tilde{E}_{H,sk,\text{III}} = -\frac{ES_z^0}{ev\mathcal D}\frac{\xi e}{\hbar l}\frac{\lambda_1^s(x)}{l}\nu\left({4+}\frac{\pi U_vNl}{v\hbar\nu}\right),
\end{equation}
described by the function $\lambda_1^s(x)$. The corresponding coordinate dependencies of the Hall voltage $V_H$, presented in Fig.~\ref{fig:Hall anomalous}(d,e), deviate from linear $l\gtrsim w^2/l_{ee}$. In the opposite limit they become linear, as expected in the diffusion regime. In the limit $l\to 0$ (but within the applicability of the kinetic equation, i.e. when $mvl/\hbar \gg 1$) the first contribution to \eqref{AHE: modB sk} from coherent scattering on impurity pairs reduces the contributions from anomalous velocities and anomalous distribution \eqref{AHE: modB v_a+v_sj+adist}. This is a feature of scattering on a short-range potential, see~\cite{Ado_2015,2020arXiv200405091G,Glazov2020b}. As a result, only the contribution from third-order asymmetric scattering with respect to the impurity potential remains, which in the deeply diffuse limit $l \ll l_{ee} \ll w$ have ceases to depend on $l$:
\begin{equation}
\label{AHE:mod:B:diff}
    \tilde{E}_{H,a}^{d} = \frac{E S_z^0 }{ev\mathcal D}\frac{\xi eE}{\hbar l}\cdot \frac{\pi U_v N}{v\hbar}.
\end{equation}

It is interesting to note that when passing from $l\rightarrow 0$ to $l\rightarrow \infty$ for a fixed $l_{ee}$ and the condition
\[
     \frac{\pi U_v N w^2}{8v\hbar l_{ee}} < 1,
\]
the direction of the Hall field can change as a function of the $x$ coordinate. In this case, the dependence of the voltage $V_{H,a}$, caused by the anomalous Hall effect, on the coordinate becomes non-monotonic, as shown in Fig.~\ref{fig:AHE sign transition} in a certain range of ratios $l/l_{ee}$. This is due to the competition between the contributions of third-order asymmetric scattering (the second term in Eq.~\eqref{AHE: modB sk}) and the anomalous contributions, Eq.~\eqref{AHE: modB v_a+v_sj+adist}. We emphasize that in narrow edge strips (marked in gray in Fig.~\ref{fig:AHE sign transition}), the expansion of the distribution function over three angular harmonics turns out to be insufficient. A full analysis of the role of such edge regions is beyond the scope of this article; preliminary estimates show that these narrow stripes do not really make a noticeable contribution to the anomalous Hall effect.

Thus, the observation of a nonlinear and, moreover, nonmonotonic dependence of the anomalous contribution to the Hall voltage on the coordinate can serve as evidence of a hydrodynamic regime of electron transport, with suppressed collisions of electrons with opposite spins (moderate magnetic fields).

\section{Conclusion}\label{sec:conclusion}

To conclude, we developed a theory of the anomalous Hall effect for two-dimensional electron gas in the case of transition between the diffusive and hydrodynamic transport regimes. All the main mechanisms of the anomalous Hall effect are taken into account: anomalous velocity, the effect of accumulation of wave packet shifts, the contribution of the anomalous distribution and the contribution from the skew scattering both on single impurities and on pairs of impurities. All these mechanisms make, generally speaking, comparable contributions to the anomalous Hall field and voltage. In the case of low magnetic fields, where the Zeeman splitting of the electron spectrum is much smaller than the thermal energy of electrons $T$, all anomalous contributions lead to a coordinate-independent Hall field and a linear coordinate dependence of the Hall voltage for any ratio between the scattering lengths on impurities $l$ and interelectronic collisions $l_{ee}$. On the contrary, the contribution of asymmetric scattering to the Hall field and voltage has a nontrivial coordinate dependence at $l_{ee} \lesssim l$ and moderate magnetic fields, where the Zeeman splitting exceeds temperature.
In the case of both low and moderate magnetic fields, the sign of the potential difference at the edges of the sample, due to anomalous contributions, can be different depending on the relationships between the parameters of the system. The effects considered are also preserved in the general case of nonparabolic electron dispersion, but specific expressions for the contributions of various mechanisms to the anomalous Hall effect differ.

Experimentally, anomalous contributions to the Hall voltage and field can be identified on top of the normal Hall effect in samples containing magnetic impurities, where the equilibrium spin polarization turns out to be a nonlinear function of the external field, or using the electron paramagnetic resonance method: By applying a weak alternating magnetic field in the channel plane with a frequency $|g\mu_B B|/\hbar$, it is possible to depolarize the charge carriers and, thereby, eliminate the anomalous contribution to the Hall effect.

\section*{Acknowledgements}

This work has been supported by the RSF project 22-12-00211. We are grateful to K.K. Grigoryan for valuable discussions.


\begin{thebibliography}{55}%
\makeatletter
\providecommand \@ifxundefined [1]{%
 \@ifx{#1\undefined}
}%
\providecommand \@ifnum [1]{%
 \ifnum #1\expandafter \@firstoftwo
 \else \expandafter \@secondoftwo
 \fi
}%
\providecommand \@ifx [1]{%
 \ifx #1\expandafter \@firstoftwo
 \else \expandafter \@secondoftwo
 \fi
}%
\providecommand \natexlab [1]{#1}%
\providecommand \enquote  [1]{``#1''}%
\providecommand \bibnamefont  [1]{#1}%
\providecommand \bibfnamefont [1]{#1}%
\providecommand \citenamefont [1]{#1}%
\providecommand \href@noop [0]{\@secondoftwo}%
\providecommand \href [0]{\begingroup \@sanitize@url \@href}%
\providecommand \@href[1]{\@@startlink{#1}\@@href}%
\providecommand \@@href[1]{\endgroup#1\@@endlink}%
\providecommand \@sanitize@url [0]{\catcode `\\12\catcode `\$12\catcode
  `\&12\catcode `\#12\catcode `\^12\catcode `\_12\catcode `\%12\relax}%
\providecommand \@@startlink[1]{}%
\providecommand \@@endlink[0]{}%
\providecommand \url  [0]{\begingroup\@sanitize@url \@url }%
\providecommand \@url [1]{\endgroup\@href {#1}{\urlprefix }}%
\providecommand \urlprefix  [0]{URL }%
\providecommand \Eprint [0]{\href }%
\providecommand \doibase [0]{http://dx.doi.org/}%
\providecommand \selectlanguage [0]{\@gobble}%
\providecommand \bibinfo  [0]{\@secondoftwo}%
\providecommand \bibfield  [0]{\@secondoftwo}%
\providecommand \translation [1]{[#1]}%
\providecommand \BibitemOpen [0]{}%
\providecommand \bibitemStop [0]{}%
\providecommand \bibitemNoStop [0]{.\EOS\space}%
\providecommand \EOS [0]{\spacefactor3000\relax}%
\providecommand \BibitemShut  [1]{\csname bibitem#1\endcsname}%
\let\auto@bib@innerbib\@empty
\bibitem [{\citenamefont {Dyakonov}(2017)}]{dyakonov_book}%
  \BibitemOpen
  \bibinfo {editor} {\bibfnamefont {M.~I.}\ \bibnamefont {Dyakonov}},\ ed.,\
  \href@noop {} {\emph {\bibinfo {title} {Spin physics in semiconductors}}},\
  \bibinfo {edition} {2nd}\ ed.,\ Springer Series in Solid-State Sciences 157\
  (\bibinfo  {publisher} {Springer International Publishing},\ \bibinfo {year}
  {2017})\BibitemShut {NoStop}%
\bibitem [{\citenamefont {Nagaosa}\ \emph {et~al.}(2010)\citenamefont
  {Nagaosa}, \citenamefont {Sinova}, \citenamefont {Onoda}, \citenamefont
  {MacDonald},\ and\ \citenamefont {Ong}}]{RevModPhys.82.1539}%
  \BibitemOpen
  \bibfield  {author} {\bibinfo {author} {\bibfnamefont {Naoto}\ \bibnamefont
  {Nagaosa}}, \bibinfo {author} {\bibfnamefont {Jairo}\ \bibnamefont {Sinova}},
  \bibinfo {author} {\bibfnamefont {Shigeki}\ \bibnamefont {Onoda}}, \bibinfo
  {author} {\bibfnamefont {A.~H.}\ \bibnamefont {MacDonald}}, \ and\ \bibinfo
  {author} {\bibfnamefont {N.~P.}\ \bibnamefont {Ong}},\ }\bibfield  {title}
  {\enquote {\bibinfo {title} {{Anomalous Hall effect}},}\ }\href {\doibase
  10.1103/RevModPhys.82.1539} {\bibfield  {journal} {\bibinfo  {journal} {Rev.
  Mod. Phys.}\ }\textbf {\bibinfo {volume} {82}},\ \bibinfo {pages}
  {1539--1592} (\bibinfo {year} {2010})}\BibitemShut {NoStop}%
\bibitem [{\citenamefont {Dyakonov}\ and\ \citenamefont
  {Perel'}(1971{\natexlab{a}})}]{dyakonov71}%
  \BibitemOpen
  \bibfield  {author} {\bibinfo {author} {\bibfnamefont {M.I.}\ \bibnamefont
  {Dyakonov}}\ and\ \bibinfo {author} {\bibfnamefont {V.I}\ \bibnamefont
  {Perel'}},\ }\bibfield  {title} {\enquote {\bibinfo {title} {{Possibility of
  Orienting Electron Spins with Current}},}\ }\href@noop {} {\bibfield
  {journal} {\bibinfo  {journal} {JETP Lett.}\ }\textbf {\bibinfo {volume}
  {13}},\ \bibinfo {pages} {657} (\bibinfo {year}
  {1971}{\natexlab{a}})}\BibitemShut {NoStop}%
\bibitem [{\citenamefont {Dyakonov}\ and\ \citenamefont
  {Perel'}(1971{\natexlab{b}})}]{dyakonov71a}%
  \BibitemOpen
  \bibfield  {author} {\bibinfo {author} {\bibfnamefont {M.~I.}\ \bibnamefont
  {Dyakonov}}\ and\ \bibinfo {author} {\bibfnamefont {V.~I.}\ \bibnamefont
  {Perel'}},\ }\bibfield  {title} {\enquote {\bibinfo {title} {Current induced
  spin orientation of electrons in semiconductors},}\ }\href
  {https://www.sciencedirect.com/science/article/pii/0375960171901964}
  {\bibfield  {journal} {\bibinfo  {journal} {Phys. Lett. A}\ }\textbf
  {\bibinfo {volume} {35A}},\ \bibinfo {pages} {459} (\bibinfo {year}
  {1971}{\natexlab{b}})}\BibitemShut {NoStop}%
\bibitem [{\citenamefont {Kato}\ \emph {et~al.}(2004)\citenamefont {Kato},
  \citenamefont {Myers}, \citenamefont {Gossard},\ and\ \citenamefont
  {Awschalom}}]{kato04}%
  \BibitemOpen
  \bibfield  {author} {\bibinfo {author} {\bibfnamefont {Y.~K.}\ \bibnamefont
  {Kato}}, \bibinfo {author} {\bibfnamefont {R.~C.}\ \bibnamefont {Myers}},
  \bibinfo {author} {\bibfnamefont {A.~C.}\ \bibnamefont {Gossard}}, \ and\
  \bibinfo {author} {\bibfnamefont {D.~D.}\ \bibnamefont {Awschalom}},\
  }\bibfield  {title} {\enquote {\bibinfo {title} {Observation of the spin
  $\mbox{H}$all effect in semiconductors},}\ }\href
  {https://science.sciencemag.org/content/306/5703/1910} {\bibfield  {journal}
  {\bibinfo  {journal} {Science}\ }\textbf {\bibinfo {volume} {306}},\ \bibinfo
  {pages} {1910} (\bibinfo {year} {2004})}\BibitemShut {NoStop}%
\bibitem [{\citenamefont {Wunderlich}\ \emph {et~al.}(2005)\citenamefont
  {Wunderlich}, \citenamefont {Kaestner}, \citenamefont {Sinova},\ and\
  \citenamefont {Jungwirth}}]{wunderlich05}%
  \BibitemOpen
  \bibfield  {author} {\bibinfo {author} {\bibfnamefont {J.}~\bibnamefont
  {Wunderlich}}, \bibinfo {author} {\bibfnamefont {B.}~\bibnamefont
  {Kaestner}}, \bibinfo {author} {\bibfnamefont {J.}~\bibnamefont {Sinova}}, \
  and\ \bibinfo {author} {\bibfnamefont {T.}~\bibnamefont {Jungwirth}},\
  }\bibfield  {title} {\enquote {\bibinfo {title} {Experimental observation of
  the spin-{H}all effect in a two-dimensional spin-orbit coupled semiconductor
  system},}\ }\href
  {https://journals.aps.org/prl/abstract/10.1103/PhysRevLett.94.047204}
  {\bibfield  {journal} {\bibinfo  {journal} {Phys. Rev. Lett.}\ }\textbf
  {\bibinfo {volume} {94}},\ \bibinfo {pages} {47204} (\bibinfo {year}
  {2005})}\BibitemShut {NoStop}%
\bibitem [{\citenamefont {Xiao}\ \emph {et~al.}(2012)\citenamefont {Xiao},
  \citenamefont {Liu}, \citenamefont {Feng}, \citenamefont {Xu},\ and\
  \citenamefont {Yao}}]{Xiao:2012cr}%
  \BibitemOpen
  \bibfield  {author} {\bibinfo {author} {\bibfnamefont {Di}~\bibnamefont
  {Xiao}}, \bibinfo {author} {\bibfnamefont {Gui-Bin}\ \bibnamefont {Liu}},
  \bibinfo {author} {\bibfnamefont {Wanxiang}\ \bibnamefont {Feng}}, \bibinfo
  {author} {\bibfnamefont {Xiaodong}\ \bibnamefont {Xu}}, \ and\ \bibinfo
  {author} {\bibfnamefont {Wang}\ \bibnamefont {Yao}},\ }\bibfield  {title}
  {\enquote {\bibinfo {title} {Coupled spin and valley physics in monolayers of
  {MoS}$_{2}$ and other group-{VI} dichalcogenides},}\ }\href {\doibase
  10.1103/PhysRevLett.108.196802} {\bibfield  {journal} {\bibinfo  {journal}
  {Phys. Rev. Lett.}\ }\textbf {\bibinfo {volume} {108}},\ \bibinfo {pages}
  {196802} (\bibinfo {year} {2012})}\BibitemShut {NoStop}%
\bibitem [{\citenamefont {Glazov}\ and\ \citenamefont
  {Golub}(2020{\natexlab{a}})}]{2020arXiv200405091G}%
  \BibitemOpen
  \bibfield  {author} {\bibinfo {author} {\bibfnamefont {M.~M.}\ \bibnamefont
  {Glazov}}\ and\ \bibinfo {author} {\bibfnamefont {L.~E.}\ \bibnamefont
  {Golub}},\ }\bibfield  {title} {\enquote {\bibinfo {title} {Valley {H}all
  effect caused by the phonon and photon drag},}\ }\href {\doibase
  10.1103/PhysRevB.102.155302} {\bibfield  {journal} {\bibinfo  {journal}
  {Phys. Rev. B}\ }\textbf {\bibinfo {volume} {102}},\ \bibinfo {pages}
  {155302} (\bibinfo {year} {2020}{\natexlab{a}})}\BibitemShut {NoStop}%
\bibitem [{\citenamefont {Hall}(1881)}]{Hall:1881aa}%
  \BibitemOpen
  \bibfield  {author} {\bibinfo {author} {\bibfnamefont {E.~H.}\ \bibnamefont
  {Hall}},\ }\bibfield  {title} {\enquote {\bibinfo {title} {{XXXVIII.} {O}n
  the new action of magnetism on a permanent electric current},}\ }\href
  {\doibase 10.1080/14786448008626936} {\bibfield  {journal} {\bibinfo
  {journal} {The London, Edinburgh, and Dublin Philosophical Magazine and
  Journal of Science}\ }\textbf {\bibinfo {volume} {5}},\ \bibinfo {pages}
  {157} (\bibinfo {year} {1881})}\BibitemShut {NoStop}%
\bibitem [{\citenamefont {Karplus}\ and\ \citenamefont
  {Luttinger}(1954)}]{PhysRev.95.1154}%
  \BibitemOpen
  \bibfield  {author} {\bibinfo {author} {\bibfnamefont {Robert}\ \bibnamefont
  {Karplus}}\ and\ \bibinfo {author} {\bibfnamefont {J.~M.}\ \bibnamefont
  {Luttinger}},\ }\bibfield  {title} {\enquote {\bibinfo {title} {Hall effect
  in ferromagnetics},}\ }\href {\doibase 10.1103/PhysRev.95.1154} {\bibfield
  {journal} {\bibinfo  {journal} {Phys. Rev.}\ }\textbf {\bibinfo {volume}
  {95}},\ \bibinfo {pages} {1154--1160} (\bibinfo {year} {1954})}\BibitemShut
  {NoStop}%
\bibitem [{\citenamefont {Smit}(1955)}]{SMIT1955877}%
  \BibitemOpen
  \bibfield  {author} {\bibinfo {author} {\bibfnamefont {J.}~\bibnamefont
  {Smit}},\ }\bibfield  {title} {\enquote {\bibinfo {title} {The spontaneous
  {H}all effect in ferromagnetics {I}},}\ }\href {\doibase
  https://doi.org/10.1016/S0031-8914(55)92596-9} {\bibfield  {journal}
  {\bibinfo  {journal} {Physica}\ }\textbf {\bibinfo {volume} {21}},\ \bibinfo
  {pages} {877 -- 887} (\bibinfo {year} {1955})}\BibitemShut {NoStop}%
\bibitem [{\citenamefont {Adams}\ and\ \citenamefont
  {Blount}(1959)}]{Adams:1959aa}%
  \BibitemOpen
  \bibfield  {author} {\bibinfo {author} {\bibfnamefont {E.~N.}\ \bibnamefont
  {Adams}}\ and\ \bibinfo {author} {\bibfnamefont {E.~I.}\ \bibnamefont
  {Blount}},\ }\bibfield  {title} {\enquote {\bibinfo {title} {Energy bands in
  the presence of an external force field---{II: A}nomalous velocities},}\
  }\href {\doibase https://doi.org/10.1016/0022-3697(59)90004-6} {\bibfield
  {journal} {\bibinfo  {journal} {Journal of Physics and Chemistry of Solids}\
  }\textbf {\bibinfo {volume} {10}},\ \bibinfo {pages} {286--303} (\bibinfo
  {year} {1959})}\BibitemShut {NoStop}%
\bibitem [{\citenamefont {Gurevich}\ and\ \citenamefont
  {Yassievich}(1963)}]{gy61}%
  \BibitemOpen
  \bibfield  {author} {\bibinfo {author} {\bibfnamefont {L.~E.}\ \bibnamefont
  {Gurevich}}\ and\ \bibinfo {author} {\bibfnamefont {I.~N.}\ \bibnamefont
  {Yassievich}},\ }\bibfield  {title} {\enquote {\bibinfo {title} {Theory of
  ferromagnetic {Hall} effect},}\ }\href@noop {} {\bibfield  {journal}
  {\bibinfo  {journal} {Sov. Phys. Solid. State}\ }\textbf {\bibinfo {volume}
  {4}},\ \bibinfo {pages} {2091} (\bibinfo {year} {1963})}\BibitemShut
  {NoStop}%
\bibitem [{\citenamefont {Abakumov}\ and\ \citenamefont
  {Yassievich}(1972)}]{abakumov72}%
  \BibitemOpen
  \bibfield  {author} {\bibinfo {author} {\bibfnamefont {V.N.}\ \bibnamefont
  {Abakumov}}\ and\ \bibinfo {author} {\bibfnamefont {I.N.}\ \bibnamefont
  {Yassievich}},\ }\bibfield  {title} {\enquote {\bibinfo {title} {Anomalous
  {H}all effect for polarized electrons in semiconductors},}\ }\href@noop {}
  {\bibfield  {journal} {\bibinfo  {journal} {JETP}\ }\textbf {\bibinfo
  {volume} {34}},\ \bibinfo {pages} {1375} (\bibinfo {year}
  {1972})}\BibitemShut {NoStop}%
\bibitem [{\citenamefont {{Nozi\`eres, P.}}\ and\ \citenamefont {{Lewiner,
  C.}}(1973)}]{nozieresAHE}%
  \BibitemOpen
  \bibfield  {author} {\bibinfo {author} {\bibnamefont {{P. Nozi\`eres,}}}\
  and\ \bibinfo {author} {\bibnamefont {{C. Lewiner}}},\ }\bibfield  {title}
  {\enquote {\bibinfo {title} {A simple theory of the anomalous {H}all effect
  in semiconductors},}\ }\href {\doibase 10.1051/jphys:019730034010090100}
  {\bibfield  {journal} {\bibinfo  {journal} {J. Phys. France}\ }\textbf
  {\bibinfo {volume} {34}},\ \bibinfo {pages} {901--915} (\bibinfo {year}
  {1973})}\BibitemShut {NoStop}%
\bibitem [{\citenamefont {Sinitsyn}\ \emph {et~al.}(2007)\citenamefont
  {Sinitsyn}, \citenamefont {MacDonald}, \citenamefont {Jungwirth},
  \citenamefont {Dugaev},\ and\ \citenamefont {Sinova}}]{PhysRevB.75.045315}%
  \BibitemOpen
  \bibfield  {author} {\bibinfo {author} {\bibfnamefont {N.~A.}\ \bibnamefont
  {Sinitsyn}}, \bibinfo {author} {\bibfnamefont {A.~H.}\ \bibnamefont
  {MacDonald}}, \bibinfo {author} {\bibfnamefont {T.}~\bibnamefont
  {Jungwirth}}, \bibinfo {author} {\bibfnamefont {V.~K.}\ \bibnamefont
  {Dugaev}}, \ and\ \bibinfo {author} {\bibfnamefont {Jairo}\ \bibnamefont
  {Sinova}},\ }\bibfield  {title} {\enquote {\bibinfo {title} {Anomalous {H}all
  effect in a two-dimensional {D}irac band: {T}he link between the
  {K}ubo-{S}treda formula and the semiclassical {B}oltzmann equation
  approach},}\ }\href {\doibase 10.1103/PhysRevB.75.045315} {\bibfield
  {journal} {\bibinfo  {journal} {Phys. Rev. B}\ }\textbf {\bibinfo {volume}
  {75}},\ \bibinfo {pages} {045315} (\bibinfo {year} {2007})}\BibitemShut
  {NoStop}%
\bibitem [{\citenamefont {Sinitsyn}(2007)}]{Sinitsyn_2007}%
  \BibitemOpen
  \bibfield  {author} {\bibinfo {author} {\bibfnamefont {N.~A.}\ \bibnamefont
  {Sinitsyn}},\ }\bibfield  {title} {\enquote {\bibinfo {title} {Semiclassical
  theories of the anomalous {H}all effect},}\ }\href {\doibase
  10.1088/0953-8984/20/02/023201} {\bibfield  {journal} {\bibinfo  {journal}
  {Journal of Physics: Condensed Matter}\ }\textbf {\bibinfo {volume} {20}},\
  \bibinfo {pages} {023201} (\bibinfo {year} {2007})}\BibitemShut {NoStop}%
\bibitem [{\citenamefont {Ado}\ \emph {et~al.}(2015)\citenamefont {Ado},
  \citenamefont {Dmitriev}, \citenamefont {Ostrovsky},\ and\ \citenamefont
  {Titov}}]{Ado_2015}%
  \BibitemOpen
  \bibfield  {author} {\bibinfo {author} {\bibfnamefont {I.~A.}\ \bibnamefont
  {Ado}}, \bibinfo {author} {\bibfnamefont {I.~A.}\ \bibnamefont {Dmitriev}},
  \bibinfo {author} {\bibfnamefont {P.~M.}\ \bibnamefont {Ostrovsky}}, \ and\
  \bibinfo {author} {\bibfnamefont {M.}~\bibnamefont {Titov}},\ }\bibfield
  {title} {\enquote {\bibinfo {title} {Anomalous {H}all effect with massive
  {D}irac fermions},}\ }\href {\doibase 10.1209/0295-5075/111/37004} {\bibfield
   {journal} {\bibinfo  {journal} {{EPL}}\ }\textbf {\bibinfo {volume} {111}},\
  \bibinfo {pages} {37004} (\bibinfo {year} {2015})}\BibitemShut {NoStop}%
\bibitem [{\citenamefont {Keser}\ \emph {et~al.}(2019)\citenamefont {Keser},
  \citenamefont {Raimondi},\ and\ \citenamefont
  {Culcer}}]{PhysRevLett.123.126603}%
  \BibitemOpen
  \bibfield  {author} {\bibinfo {author} {\bibfnamefont {Ayd\ifmmode \imath
  \else \i \fi{}n~Cem}\ \bibnamefont {Keser}}, \bibinfo {author} {\bibfnamefont
  {Roberto}\ \bibnamefont {Raimondi}}, \ and\ \bibinfo {author} {\bibfnamefont
  {Dimitrie}\ \bibnamefont {Culcer}},\ }\bibfield  {title} {\enquote {\bibinfo
  {title} {Sign change in the anomalous {H}all effect and strong transport
  effects in a {2D} massive {D}irac metal due to spin-charge correlated
  disorder},}\ }\href {\doibase 10.1103/PhysRevLett.123.126603} {\bibfield
  {journal} {\bibinfo  {journal} {Phys. Rev. Lett.}\ }\textbf {\bibinfo
  {volume} {123}},\ \bibinfo {pages} {126603} (\bibinfo {year}
  {2019})}\BibitemShut {NoStop}%
\bibitem [{\citenamefont {Belinicher}\ \emph {et~al.}(1982)\citenamefont
  {Belinicher}, \citenamefont {Ivchenko},\ and\ \citenamefont
  {Sturman}}]{belinicher82}%
  \BibitemOpen
  \bibfield  {author} {\bibinfo {author} {\bibfnamefont {V.~I.}\ \bibnamefont
  {Belinicher}}, \bibinfo {author} {\bibfnamefont {E.~L.}\ \bibnamefont
  {Ivchenko}}, \ and\ \bibinfo {author} {\bibfnamefont {B.~I.}\ \bibnamefont
  {Sturman}},\ }\bibfield  {title} {\enquote {\bibinfo {title} {Kinetic theory
  of the displacement photovoltaic effect in piezoelectrics},}\ }\href
  {http://www.jetp.ac.ru/cgi-bin/e/index/e/56/2/p359?a=list} {\bibfield
  {journal} {\bibinfo  {journal} {JETP}\ }\textbf {\bibinfo {volume} {56}},\
  \bibinfo {pages} {359} (\bibinfo {year} {1982})}\BibitemShut {NoStop}%
\bibitem [{\citenamefont {de~Jong}\ and\ \citenamefont
  {Molenkamp}(1995)}]{PhysRevB.51.13389}%
  \BibitemOpen
  \bibfield  {author} {\bibinfo {author} {\bibfnamefont {M.~J.~M.}\
  \bibnamefont {de~Jong}}\ and\ \bibinfo {author} {\bibfnamefont {L.~W.}\
  \bibnamefont {Molenkamp}},\ }\bibfield  {title} {\enquote {\bibinfo {title}
  {Hydrodynamic electron flow in high-mobility wires},}\ }\href {\doibase
  10.1103/PhysRevB.51.13389} {\bibfield  {journal} {\bibinfo  {journal} {Phys.
  Rev. B}\ }\textbf {\bibinfo {volume} {51}},\ \bibinfo {pages} {13389--13402}
  (\bibinfo {year} {1995})}\BibitemShut {NoStop}%
\bibitem [{\citenamefont {Bandurin}\ \emph {et~al.}(2016)\citenamefont
  {Bandurin}, \citenamefont {Torre}, \citenamefont {Kumar}, \citenamefont
  {Ben~Shalom}, \citenamefont {Tomadin}, \citenamefont {Principi},
  \citenamefont {Auton}, \citenamefont {Khestanova}, \citenamefont {Novoselov},
  \citenamefont {Grigorieva}, \citenamefont {Ponomarenko}, \citenamefont
  {Geim},\ and\ \citenamefont {Polini}}]{Bandurin1055}%
  \BibitemOpen
  \bibfield  {author} {\bibinfo {author} {\bibfnamefont {D.~A.}\ \bibnamefont
  {Bandurin}}, \bibinfo {author} {\bibfnamefont {I.}~\bibnamefont {Torre}},
  \bibinfo {author} {\bibfnamefont {R.~Krishna}\ \bibnamefont {Kumar}},
  \bibinfo {author} {\bibfnamefont {M.}~\bibnamefont {Ben~Shalom}}, \bibinfo
  {author} {\bibfnamefont {A.}~\bibnamefont {Tomadin}}, \bibinfo {author}
  {\bibfnamefont {A.}~\bibnamefont {Principi}}, \bibinfo {author}
  {\bibfnamefont {G.~H.}\ \bibnamefont {Auton}}, \bibinfo {author}
  {\bibfnamefont {E.}~\bibnamefont {Khestanova}}, \bibinfo {author}
  {\bibfnamefont {K.~S.}\ \bibnamefont {Novoselov}}, \bibinfo {author}
  {\bibfnamefont {I.~V.}\ \bibnamefont {Grigorieva}}, \bibinfo {author}
  {\bibfnamefont {L.~A.}\ \bibnamefont {Ponomarenko}}, \bibinfo {author}
  {\bibfnamefont {A.~K.}\ \bibnamefont {Geim}}, \ and\ \bibinfo {author}
  {\bibfnamefont {M.}~\bibnamefont {Polini}},\ }\bibfield  {title} {\enquote
  {\bibinfo {title} {Negative local resistance caused by viscous electron
  backflow in graphene},}\ }\href {\doibase 10.1126/science.aad0201} {\bibfield
   {journal} {\bibinfo  {journal} {Science}\ }\textbf {\bibinfo {volume}
  {351}},\ \bibinfo {pages} {1055--1058} (\bibinfo {year} {2016})}\BibitemShut
  {NoStop}%
\bibitem [{\citenamefont {Crossno}\ \emph {et~al.}(2016)\citenamefont
  {Crossno}, \citenamefont {Shi}, \citenamefont {Wang}, \citenamefont {Liu},
  \citenamefont {Harzheim}, \citenamefont {Lucas}, \citenamefont {Sachdev},
  \citenamefont {Kim}, \citenamefont {Taniguchi}, \citenamefont {Watanabe},
  \citenamefont {Ohki},\ and\ \citenamefont {Fong}}]{Crossno:2016aa}%
  \BibitemOpen
  \bibfield  {author} {\bibinfo {author} {\bibfnamefont {Jesse}\ \bibnamefont
  {Crossno}}, \bibinfo {author} {\bibfnamefont {Jing~K.}\ \bibnamefont {Shi}},
  \bibinfo {author} {\bibfnamefont {Ke}~\bibnamefont {Wang}}, \bibinfo {author}
  {\bibfnamefont {Xiaomeng}\ \bibnamefont {Liu}}, \bibinfo {author}
  {\bibfnamefont {Achim}\ \bibnamefont {Harzheim}}, \bibinfo {author}
  {\bibfnamefont {Andrew}\ \bibnamefont {Lucas}}, \bibinfo {author}
  {\bibfnamefont {Subir}\ \bibnamefont {Sachdev}}, \bibinfo {author}
  {\bibfnamefont {Philip}\ \bibnamefont {Kim}}, \bibinfo {author}
  {\bibfnamefont {Takashi}\ \bibnamefont {Taniguchi}}, \bibinfo {author}
  {\bibfnamefont {Kenji}\ \bibnamefont {Watanabe}}, \bibinfo {author}
  {\bibfnamefont {Thomas~A.}\ \bibnamefont {Ohki}}, \ and\ \bibinfo {author}
  {\bibfnamefont {Kin~Chung}\ \bibnamefont {Fong}},\ }\bibfield  {title}
  {\enquote {\bibinfo {title} {{Observation of the Dirac fluid and the
  breakdown of the Wiedemann-Franz law in graphene}},}\ }\href {\doibase
  10.1126/science.aad0343} {\bibfield  {journal} {\bibinfo  {journal}
  {Science}\ }\textbf {\bibinfo {volume} {351}},\ \bibinfo {pages} {1058--1061}
  (\bibinfo {year} {2016})}\BibitemShut {NoStop}%
\bibitem [{\citenamefont {Moll}\ \emph {et~al.}(2016)\citenamefont {Moll},
  \citenamefont {Kushwaha}, \citenamefont {Nandi}, \citenamefont {Schmidt},\
  and\ \citenamefont {Mackenzie}}]{Moll1061}%
  \BibitemOpen
  \bibfield  {author} {\bibinfo {author} {\bibfnamefont {Philip J.~W.}\
  \bibnamefont {Moll}}, \bibinfo {author} {\bibfnamefont {Pallavi}\
  \bibnamefont {Kushwaha}}, \bibinfo {author} {\bibfnamefont {Nabhanila}\
  \bibnamefont {Nandi}}, \bibinfo {author} {\bibfnamefont {Burkhard}\
  \bibnamefont {Schmidt}}, \ and\ \bibinfo {author} {\bibfnamefont {Andrew~P.}\
  \bibnamefont {Mackenzie}},\ }\bibfield  {title} {\enquote {\bibinfo {title}
  {{Evidence for hydrodynamic electron flow in PdCoO$_2$}},}\ }\href {\doibase
  10.1126/science.aac8385} {\bibfield  {journal} {\bibinfo  {journal}
  {Science}\ }\textbf {\bibinfo {volume} {351}},\ \bibinfo {pages} {1061--1064}
  (\bibinfo {year} {2016})}\BibitemShut {NoStop}%
  \bibitem [{\citenamefont {Alekseev}(2016)}]{PhysRevLett.117.166601}%
  \BibitemOpen
  \bibfield  {author} {\bibinfo {author} {\bibfnamefont {P.~S.}\ \bibnamefont
  {Alekseev}},\ }\bibfield  {title} {\enquote {\bibinfo {title} {Negative
  magnetoresistance in viscous flow of two-dimensional electrons},}\ }\href
  {\doibase 10.1103/PhysRevLett.117.166601} {\bibfield  {journal} {\bibinfo
  {journal} {Phys. Rev. Lett.}\ }\textbf {\bibinfo {volume} {117}},\ \bibinfo
  {pages} {166601} (\bibinfo {year} {2016})}\BibitemShut {NoStop}%
\bibitem [{\citenamefont {Krishna~Kumar}\ \emph {et~al.}(2017)\citenamefont
  {Krishna~Kumar}, \citenamefont {Bandurin}, \citenamefont {Pellegrino},
  \citenamefont {Cao}, \citenamefont {Principi}, \citenamefont {Guo},
  \citenamefont {Auton}, \citenamefont {Ben~Shalom}, \citenamefont
  {Ponomarenko}, \citenamefont {Falkovich}, \citenamefont {Watanabe},
  \citenamefont {Taniguchi}, \citenamefont {Grigorieva}, \citenamefont
  {Levitov}, \citenamefont {Polini},\ and\ \citenamefont
  {Geim}}]{Krishna-Kumar:2017wn}%
  \BibitemOpen
  \bibfield  {author} {\bibinfo {author} {\bibfnamefont {R.}~\bibnamefont
  {Krishna~Kumar}}, \bibinfo {author} {\bibfnamefont {D.~A.}\ \bibnamefont
  {Bandurin}}, \bibinfo {author} {\bibfnamefont {F.~M.~D.}\ \bibnamefont
  {Pellegrino}}, \bibinfo {author} {\bibfnamefont {Y.}~\bibnamefont {Cao}},
  \bibinfo {author} {\bibfnamefont {A.}~\bibnamefont {Principi}}, \bibinfo
  {author} {\bibfnamefont {H.}~\bibnamefont {Guo}}, \bibinfo {author}
  {\bibfnamefont {G.~H.}\ \bibnamefont {Auton}}, \bibinfo {author}
  {\bibfnamefont {M.}~\bibnamefont {Ben~Shalom}}, \bibinfo {author}
  {\bibfnamefont {L.~A.}\ \bibnamefont {Ponomarenko}}, \bibinfo {author}
  {\bibfnamefont {G.}~\bibnamefont {Falkovich}}, \bibinfo {author}
  {\bibfnamefont {K.}~\bibnamefont {Watanabe}}, \bibinfo {author}
  {\bibfnamefont {T.}~\bibnamefont {Taniguchi}}, \bibinfo {author}
  {\bibfnamefont {I.~V.}\ \bibnamefont {Grigorieva}}, \bibinfo {author}
  {\bibfnamefont {L.~S.}\ \bibnamefont {Levitov}}, \bibinfo {author}
  {\bibfnamefont {M.}~\bibnamefont {Polini}}, \ and\ \bibinfo {author}
  {\bibfnamefont {A.~K.}\ \bibnamefont {Geim}},\ }\bibfield  {title} {\enquote
  {\bibinfo {title} {Superballistic flow of viscous electron fluid through
  graphene constrictions},}\ }\href {\doibase 10.1038/nphys4240} {\bibfield
  {journal} {\bibinfo  {journal} {Nature Physics}\ }\textbf {\bibinfo {volume}
  {13}},\ \bibinfo {pages} {1182--1185} (\bibinfo {year} {2017})}\BibitemShut
  {NoStop}%
\bibitem [{\citenamefont {Gusev}\ \emph {et~al.}(2018)\citenamefont {Gusev},
  \citenamefont {Levin}, \citenamefont {Levinson},\ and\ \citenamefont
  {Bakarov}}]{Gusev:2018tg}%
  \BibitemOpen
  \bibfield  {author} {\bibinfo {author} {\bibfnamefont {G.~M.}\ \bibnamefont
  {Gusev}}, \bibinfo {author} {\bibfnamefont {A.~D.}\ \bibnamefont {Levin}},
  \bibinfo {author} {\bibfnamefont {E.~V.}\ \bibnamefont {Levinson}}, \ and\
  \bibinfo {author} {\bibfnamefont {A.~K.}\ \bibnamefont {Bakarov}},\
  }\bibfield  {title} {\enquote {\bibinfo {title} {Viscous electron flow in
  mesoscopic two-dimensional electron gas},}\ }\href {\doibase
  10.1063/1.5020763} {\bibfield  {journal} {\bibinfo  {journal} {AIP Advances}\
  }\textbf {\bibinfo {volume} {8}},\ \bibinfo {pages} {025318} (\bibinfo {year}
  {2018})}\BibitemShut {NoStop}%
\bibitem [{\citenamefont {Pusep}\ \emph {et~al.}(2022)\citenamefont {Pusep},
  \citenamefont {Teodoro}, \citenamefont {Laurindo}, \citenamefont {Cardozo~de
  Oliveira}, \citenamefont {Gusev},\ and\ \citenamefont
  {Bakarov}}]{PhysRevLett.128.136801}%
  \BibitemOpen
  \bibfield  {author} {\bibinfo {author} {\bibfnamefont {Yu.~A.}\ \bibnamefont
  {Pusep}}, \bibinfo {author} {\bibfnamefont {M.~D.}\ \bibnamefont {Teodoro}},
  \bibinfo {author} {\bibfnamefont {V.}~\bibnamefont {Laurindo}}, \bibinfo
  {author} {\bibfnamefont {E.~R.}\ \bibnamefont {Cardozo~de Oliveira}},
  \bibinfo {author} {\bibfnamefont {G.~M.}\ \bibnamefont {Gusev}}, \ and\
  \bibinfo {author} {\bibfnamefont {A.~K.}\ \bibnamefont {Bakarov}},\
  }\bibfield  {title} {\enquote {\bibinfo {title} {Diffusion of photoexcited
  holes in a viscous electron fluid},}\ }\href {\doibase
  10.1103/PhysRevLett.128.136801} {\bibfield  {journal} {\bibinfo  {journal}
  {Phys. Rev. Lett.}\ }\textbf {\bibinfo {volume} {128}},\ \bibinfo {pages}
  {136801} (\bibinfo {year} {2022})}\BibitemShut {NoStop}%
\bibitem [{\citenamefont {Gurzhi}(1963)}]{gurzhi63}%
  \BibitemOpen
  \bibfield  {author} {\bibinfo {author} {\bibfnamefont {R.~N.}\ \bibnamefont
  {Gurzhi}},\ }\bibfield  {title} {\enquote {\bibinfo {title} {Minimum of
  resistance in impurity-free conductors},}\ }\href@noop {} {\bibfield
  {journal} {\bibinfo  {journal} {JETP}\ }\textbf {\bibinfo {volume} {17}},\
  \bibinfo {pages} {521} (\bibinfo {year} {1963})}\BibitemShut {NoStop}%
\bibitem [{\citenamefont {Gurzhi}(1968)}]{Gurzhi_1968}%
  \BibitemOpen
  \bibfield  {author} {\bibinfo {author} {\bibfnamefont {R.~N.}\ \bibnamefont
  {Gurzhi}},\ }\bibfield  {title} {\enquote {\bibinfo {title} {Hydrodynamic
  effects in solids at low temperatures},}\ }\href {\doibase
  10.1070/pu1968v011n02abeh003815} {\bibfield  {journal} {\bibinfo  {journal}
  {Soviet Physics Uspekhi}\ }\textbf {\bibinfo {volume} {11}},\ \bibinfo
  {pages} {255--270} (\bibinfo {year} {1968})}\BibitemShut {NoStop}%
\bibitem [{\citenamefont {Andreev}\ \emph {et~al.}(2011)\citenamefont
  {Andreev}, \citenamefont {Kivelson},\ and\ \citenamefont
  {Spivak}}]{PhysRevLett.106.256804}%
  \BibitemOpen
  \bibfield  {author} {\bibinfo {author} {\bibfnamefont {A.~V.}\ \bibnamefont
  {Andreev}}, \bibinfo {author} {\bibfnamefont {Steven~A.}\ \bibnamefont
  {Kivelson}}, \ and\ \bibinfo {author} {\bibfnamefont {B.}~\bibnamefont
  {Spivak}},\ }\bibfield  {title} {\enquote {\bibinfo {title} {Hydrodynamic
  description of transport in strongly correlated electron systems},}\ }\href
  {\doibase 10.1103/PhysRevLett.106.256804} {\bibfield  {journal} {\bibinfo
  {journal} {Phys. Rev. Lett.}\ }\textbf {\bibinfo {volume} {106}},\ \bibinfo
  {pages} {256804} (\bibinfo {year} {2011})}\BibitemShut {NoStop}%
\bibitem [{\citenamefont {Torre}\ \emph {et~al.}(2015)\citenamefont {Torre},
  \citenamefont {Tomadin}, \citenamefont {Geim},\ and\ \citenamefont
  {Polini}}]{PhysRevB.92.165433}%
  \BibitemOpen
  \bibfield  {author} {\bibinfo {author} {\bibfnamefont {Iacopo}\ \bibnamefont
  {Torre}}, \bibinfo {author} {\bibfnamefont {Andrea}\ \bibnamefont {Tomadin}},
  \bibinfo {author} {\bibfnamefont {Andre~K.}\ \bibnamefont {Geim}}, \ and\
  \bibinfo {author} {\bibfnamefont {Marco}\ \bibnamefont {Polini}},\ }\bibfield
   {title} {\enquote {\bibinfo {title} {Nonlocal transport and the hydrodynamic
  shear viscosity in graphene},}\ }\href {\doibase 10.1103/PhysRevB.92.165433}
  {\bibfield  {journal} {\bibinfo  {journal} {Phys. Rev. B}\ }\textbf {\bibinfo
  {volume} {92}},\ \bibinfo {pages} {165433} (\bibinfo {year}
  {2015})}\BibitemShut {NoStop}%
\bibitem [{\citenamefont {Levitov}\ and\ \citenamefont
  {Falkovich}(2016)}]{Levitov:2016aa}%
  \BibitemOpen
  \bibfield  {author} {\bibinfo {author} {\bibfnamefont {Leonid}\ \bibnamefont
  {Levitov}}\ and\ \bibinfo {author} {\bibfnamefont {Gregory}\ \bibnamefont
  {Falkovich}},\ }\bibfield  {title} {\enquote {\bibinfo {title} {Electron
  viscosity, current vortices and negative nonlocal resistance in graphene},}\
  }\href {\doibase 10.1038/nphys3667} {\bibfield  {journal} {\bibinfo
  {journal} {Nature Physics}\ }\textbf {\bibinfo {volume} {12}},\ \bibinfo
  {pages} {672--676} (\bibinfo {year} {2016})}\BibitemShut {NoStop}%
\bibitem [{\citenamefont {Scaffidi}\ \emph {et~al.}(2017)\citenamefont
  {Scaffidi}, \citenamefont {Nandi}, \citenamefont {Schmidt}, \citenamefont
  {Mackenzie},\ and\ \citenamefont {Moore}}]{PhysRevLett.118.226601}%
  \BibitemOpen
  \bibfield  {author} {\bibinfo {author} {\bibfnamefont {Thomas}\ \bibnamefont
  {Scaffidi}}, \bibinfo {author} {\bibfnamefont {Nabhanila}\ \bibnamefont
  {Nandi}}, \bibinfo {author} {\bibfnamefont {Burkhard}\ \bibnamefont
  {Schmidt}}, \bibinfo {author} {\bibfnamefont {Andrew~P.}\ \bibnamefont
  {Mackenzie}}, \ and\ \bibinfo {author} {\bibfnamefont {Joel~E.}\ \bibnamefont
  {Moore}},\ }\bibfield  {title} {\enquote {\bibinfo {title} {Hydrodynamic
  electron flow and {Hall} viscosity},}\ }\href {\doibase
  10.1103/PhysRevLett.118.226601} {\bibfield  {journal} {\bibinfo  {journal}
  {Phys. Rev. Lett.}\ }\textbf {\bibinfo {volume} {118}},\ \bibinfo {pages}
  {226601} (\bibinfo {year} {2017})}\BibitemShut {NoStop}%
\bibitem [{\citenamefont {Apostolov}\ \emph {et~al.}(2019)\citenamefont
  {Apostolov}, \citenamefont {Pesin},\ and\ \citenamefont
  {Levchenko}}]{PhysRevB.100.115401}%
  \BibitemOpen
  \bibfield  {author} {\bibinfo {author} {\bibfnamefont {S.~S.}\ \bibnamefont
  {Apostolov}}, \bibinfo {author} {\bibfnamefont {D.~A.}\ \bibnamefont
  {Pesin}}, \ and\ \bibinfo {author} {\bibfnamefont {A.}~\bibnamefont
  {Levchenko}},\ }\bibfield  {title} {\enquote {\bibinfo {title} {Magnetodrag
  in the hydrodynamic regime: Effects of magnetoplasmon resonance and {H}all
  viscosity},}\ }\href {\doibase 10.1103/PhysRevB.100.115401} {\bibfield
  {journal} {\bibinfo  {journal} {Phys. Rev. B}\ }\textbf {\bibinfo {volume}
  {100}},\ \bibinfo {pages} {115401} (\bibinfo {year} {2019})}\BibitemShut
  {NoStop}%
\bibitem [{\citenamefont {Narozhny}\ \emph {et~al.}(2017)\citenamefont
  {Narozhny}, \citenamefont {Gornyi}, \citenamefont {Mirlin},\ and\
  \citenamefont {Schmalian}}]{Narozhny:2017vc}%
  \BibitemOpen
  \bibfield  {author} {\bibinfo {author} {\bibfnamefont {Boris~N.}\
  \bibnamefont {Narozhny}}, \bibinfo {author} {\bibfnamefont {Igor~V.}\
  \bibnamefont {Gornyi}}, \bibinfo {author} {\bibfnamefont {Alexander~D.}\
  \bibnamefont {Mirlin}}, \ and\ \bibinfo {author} {\bibfnamefont {J{\"o}rg}\
  \bibnamefont {Schmalian}},\ }\bibfield  {title} {\enquote {\bibinfo {title}
  {Hydrodynamic approach to electronic transport in graphene},}\ }\href
  {\doibase https://doi.org/10.1002/andp.201700043} {\bibfield  {journal}
  {\bibinfo  {journal} {Annalen der Physik}\ }\textbf {\bibinfo {volume}
  {529}},\ \bibinfo {pages} {1700043} (\bibinfo {year} {2017})}\BibitemShut
  {NoStop}%
\bibitem [{\citenamefont {Narozhny}(2022)}]{Narozhny:2022ud}%
  \BibitemOpen
  \bibfield  {author} {\bibinfo {author} {\bibfnamefont {Boris~N.}\
  \bibnamefont {Narozhny}},\ }\bibfield  {title} {\enquote {\bibinfo {title}
  {Hydrodynamic approach to two-dimensional electron systems},}\ }\href
  {\doibase 10.1007/s40766-022-00036-z} {\bibfield  {journal} {\bibinfo
  {journal} {La Rivista del Nuovo Cimento}\ }\textbf {\bibinfo {volume} {45}},\
  \bibinfo {pages} {661} (\bibinfo {year} {2022})}\BibitemShut {NoStop}%
\bibitem [{\citenamefont {Pesin}(2018)}]{PhysRevLett.121.226601}%
  \BibitemOpen
  \bibfield  {author} {\bibinfo {author} {\bibfnamefont {D.~A.}\ \bibnamefont
  {Pesin}},\ }\bibfield  {title} {\enquote {\bibinfo {title} {{Two-Particle
  Collisional Coordinate Shifts and Hydrodynamic Anomalous Hall Effect in
  Systems without Lorentz Invariance}},}\ }\href {\doibase
  10.1103/PhysRevLett.121.226601} {\bibfield  {journal} {\bibinfo  {journal}
  {Phys. Rev. Lett.}\ }\textbf {\bibinfo {volume} {121}},\ \bibinfo {pages}
  {226601} (\bibinfo {year} {2018})}\BibitemShut {NoStop}%
\bibitem [{\citenamefont {Funaki}\ \emph {et~al.}(2021)\citenamefont {Funaki},
  \citenamefont {Toshio},\ and\ \citenamefont
  {Tatara}}]{PhysRevResearch.3.033075}%
  \BibitemOpen
  \bibfield  {author} {\bibinfo {author} {\bibfnamefont {Hiroshi}\ \bibnamefont
  {Funaki}}, \bibinfo {author} {\bibfnamefont {Riki}\ \bibnamefont {Toshio}}, \
  and\ \bibinfo {author} {\bibfnamefont {Gen}\ \bibnamefont {Tatara}},\
  }\bibfield  {title} {\enquote {\bibinfo {title} {Vorticity-induced anomalous
  {Hall} effect in an electron fluid},}\ }\href {\doibase
  10.1103/PhysRevResearch.3.033075} {\bibfield  {journal} {\bibinfo  {journal}
  {Phys. Rev. Res.}\ }\textbf {\bibinfo {volume} {3}},\ \bibinfo {pages}
  {033075} (\bibinfo {year} {2021})}\BibitemShut {NoStop}%
\bibitem [{\citenamefont {Tatara}(2021)}]{PhysRevB.104.184414}%
  \BibitemOpen
  \bibfield  {author} {\bibinfo {author} {\bibfnamefont {Gen}\ \bibnamefont
  {Tatara}},\ }\bibfield  {title} {\enquote {\bibinfo {title} {Hydrodynamic
  theory of vorticity-induced spin transport},}\ }\href {\doibase
  10.1103/PhysRevB.104.184414} {\bibfield  {journal} {\bibinfo  {journal}
  {Phys. Rev. B}\ }\textbf {\bibinfo {volume} {104}},\ \bibinfo {pages}
  {184414} (\bibinfo {year} {2021})}\BibitemShut {NoStop}%
\bibitem [{\citenamefont {Hasdeo}\ \emph {et~al.}(2021)\citenamefont {Hasdeo},
  \citenamefont {Ekstr\"om}, \citenamefont {Idrisov},\ and\ \citenamefont
  {Schmidt}}]{PhysRevB.103.125106}%
  \BibitemOpen
  \bibfield  {author} {\bibinfo {author} {\bibfnamefont {Eddwi~H.}\
  \bibnamefont {Hasdeo}}, \bibinfo {author} {\bibfnamefont {Johan}\
  \bibnamefont {Ekstr\"om}}, \bibinfo {author} {\bibfnamefont {Edvin~G.}\
  \bibnamefont {Idrisov}}, \ and\ \bibinfo {author} {\bibfnamefont {Thomas~L.}\
  \bibnamefont {Schmidt}},\ }\bibfield  {title} {\enquote {\bibinfo {title}
  {Electron hydrodynamics of two-dimensional anomalous {H}all materials},}\
  }\href {\doibase 10.1103/PhysRevB.103.125106} {\bibfield  {journal} {\bibinfo
   {journal} {Phys. Rev. B}\ }\textbf {\bibinfo {volume} {103}},\ \bibinfo
  {pages} {125106} (\bibinfo {year} {2021})}\BibitemShut {NoStop}%
\bibitem [{\citenamefont {Glazov}(2022)}]{Glazov_2021b}%
  \BibitemOpen
  \bibfield  {author} {\bibinfo {author} {\bibfnamefont {M~M}\ \bibnamefont
  {Glazov}},\ }\bibfield  {title} {\enquote {\bibinfo {title} {Valley and spin
  accumulation in ballistic and hydrodynamic channels},}\ }\href {\doibase
  10.1088/2053-1583/ac3e04} {\bibfield  {journal} {\bibinfo  {journal} {2D
  Materials}\ }\textbf {\bibinfo {volume} {9}},\ \bibinfo {pages} {015027}
  (\bibinfo {year} {2022})}\BibitemShut {NoStop}%
\bibitem [{\citenamefont {Afanasiev}\ \emph
  {et~al.}(2022{\natexlab{a}})\citenamefont {Afanasiev}, \citenamefont
  {Alekseev}, \citenamefont {Danilenko}, \citenamefont {Greshnov},\ and\
  \citenamefont {Semina}}]{PhysRevB.106.L041407}%
  \BibitemOpen
  \bibfield  {author} {\bibinfo {author} {\bibfnamefont {A.~N.}\ \bibnamefont
  {Afanasiev}}, \bibinfo {author} {\bibfnamefont {P.~S.}\ \bibnamefont
  {Alekseev}}, \bibinfo {author} {\bibfnamefont {A.~A.}\ \bibnamefont
  {Danilenko}}, \bibinfo {author} {\bibfnamefont {A.~A.}\ \bibnamefont
  {Greshnov}}, \ and\ \bibinfo {author} {\bibfnamefont {M.~A.}\ \bibnamefont
  {Semina}},\ }\bibfield  {title} {\enquote {\bibinfo {title} {Rotational
  viscosity in spin resonance of hydrodynamic electrons},}\ }\href {\doibase
  10.1103/PhysRevB.106.L041407} {\bibfield  {journal} {\bibinfo  {journal}
  {Phys. Rev. B}\ }\textbf {\bibinfo {volume} {106}},\ \bibinfo {pages}
  {L041407} (\bibinfo {year} {2022}{\natexlab{a}})}\BibitemShut {NoStop}%
\bibitem [{\citenamefont {Grigoryan}\ \emph {et~al.}(2023)\citenamefont
  {Grigoryan}, \citenamefont {Zohrabyan},\ and\ \citenamefont
  {Glazov}}]{grigoryan2023anomalous}%
  \BibitemOpen
  \bibfield  {author} {\bibinfo {author} {\bibfnamefont {K.~K.}\ \bibnamefont
  {Grigoryan}}, \bibinfo {author} {\bibfnamefont {D.~S.}\ \bibnamefont
  {Zohrabyan}}, \ and\ \bibinfo {author} {\bibfnamefont {M.~M.}\ \bibnamefont
  {Glazov}},\ }\href@noop {} {\enquote {\bibinfo {title} {Anomalous Hall effect
  in ultraclean electronic channels},}\ } (\bibinfo {year} {2023}),\ \Eprint
  {http://arxiv.org/abs/2309.05401} {arXiv:2309.05401}
  \BibitemShut {NoStop}%
\bibitem [{\citenamefont {Alekseev}(2018)}]{PhysRevB.98.165440}%
  \BibitemOpen
  \bibfield  {author} {\bibinfo {author} {\bibfnamefont {P.~S.}\ \bibnamefont
  {Alekseev}},\ }\bibfield  {title} {\enquote {\bibinfo {title} {Magnetic
  resonance in a high-frequency flow of a two-dimensional viscous electron
  fluid},}\ }\href {\doibase 10.1103/PhysRevB.98.165440} {\bibfield  {journal}
  {\bibinfo  {journal} {Phys. Rev. B}\ }\textbf {\bibinfo {volume} {98}},\
  \bibinfo {pages} {165440} (\bibinfo {year} {2018})}\BibitemShut {NoStop}%
\bibitem [{\citenamefont {Alekseev}\ and\ \citenamefont
  {Semina}(2019)}]{PhysRevB.100.125419}%
  \BibitemOpen
  \bibfield  {author} {\bibinfo {author} {\bibfnamefont {P.~S.}\ \bibnamefont
  {Alekseev}}\ and\ \bibinfo {author} {\bibfnamefont {M.~A.}\ \bibnamefont
  {Semina}},\ }\bibfield  {title} {\enquote {\bibinfo {title} {Hall effect in a
  ballistic flow of two-dimensional interacting particles},}\ }\href {\doibase
  10.1103/PhysRevB.100.125419} {\bibfield  {journal} {\bibinfo  {journal}
  {Phys. Rev. B}\ }\textbf {\bibinfo {volume} {100}},\ \bibinfo {pages}
  {125419} (\bibinfo {year} {2019})}\BibitemShut {NoStop}%
\bibitem [{\citenamefont {Alekseev}\ and\ \citenamefont
  {Semina}(2018)}]{PhysRevB.98.165412}%
  \BibitemOpen
  \bibfield  {author} {\bibinfo {author} {\bibfnamefont {P.~S.}\ \bibnamefont
  {Alekseev}}\ and\ \bibinfo {author} {\bibfnamefont {M.~A.}\ \bibnamefont
  {Semina}},\ }\bibfield  {title} {\enquote {\bibinfo {title} {Ballistic flow
  of two-dimensional interacting electrons},}\ }\href {\doibase
  10.1103/PhysRevB.98.165412} {\bibfield  {journal} {\bibinfo  {journal} {Phys.
  Rev. B}\ }\textbf {\bibinfo {volume} {98}},\ \bibinfo {pages} {165412}
  (\bibinfo {year} {2018})}\BibitemShut {NoStop}%
\bibitem [{\citenamefont {Alekseev}\ and\ \citenamefont
  {Dmitriev}(2021)}]{PhysRevB.104.085434}%
  \BibitemOpen
  \bibfield  {author} {\bibinfo {author} {\bibfnamefont {Yu.~O.}\ \bibnamefont
  {Alekseev}}\ and\ \bibinfo {author} {\bibfnamefont {A.~P.}\ \bibnamefont
  {Dmitriev}},\ }\bibfield  {title} {\enquote {\bibinfo {title} {Giant {Hall}
  effect in the ballistic transport of two-dimensional electrons},}\ }\href
  {\doibase 10.1103/PhysRevB.104.085434} {\bibfield  {journal} {\bibinfo
  {journal} {Phys. Rev. B}\ }\textbf {\bibinfo {volume} {104}},\ \bibinfo
  {pages} {085434} (\bibinfo {year} {2021})}\BibitemShut {NoStop}%
\bibitem [{\citenamefont {Afanasiev}\ \emph
  {et~al.}(2022{\natexlab{b}})\citenamefont {Afanasiev}, \citenamefont
  {Alekseev}, \citenamefont {Danilenko}, \citenamefont {Dmitriev},
  \citenamefont {Greshnov},\ and\ \citenamefont
  {Semina}}]{PhysRevB.106.245415}%
  \BibitemOpen
  \bibfield  {author} {\bibinfo {author} {\bibfnamefont {A.~N.}\ \bibnamefont
  {Afanasiev}}, \bibinfo {author} {\bibfnamefont {P.~S.}\ \bibnamefont
  {Alekseev}}, \bibinfo {author} {\bibfnamefont {A.~A.}\ \bibnamefont
  {Danilenko}}, \bibinfo {author} {\bibfnamefont {A.~P.}\ \bibnamefont
  {Dmitriev}}, \bibinfo {author} {\bibfnamefont {A.~A.}\ \bibnamefont
  {Greshnov}}, \ and\ \bibinfo {author} {\bibfnamefont {M.~A.}\ \bibnamefont
  {Semina}},\ }\bibfield  {title} {\enquote {\bibinfo {title} {Hall effect in
  {Poiseuille} flow of two-dimensional electron fluid},}\ }\href {\doibase
  10.1103/PhysRevB.106.245415} {\bibfield  {journal} {\bibinfo  {journal}
  {Phys. Rev. B}\ }\textbf {\bibinfo {volume} {106}},\ \bibinfo {pages}
  {245415} (\bibinfo {year} {2022}{\natexlab{b}})}\BibitemShut {NoStop}%
\bibitem [{\citenamefont {Glazov}\ and\ \citenamefont
  {Ivchenko}(2002)}]{glazov02}%
  \BibitemOpen
  \bibfield  {author} {\bibinfo {author} {\bibfnamefont {M.~M.}\ \bibnamefont
  {Glazov}}\ and\ \bibinfo {author} {\bibfnamefont {E.~L.}\ \bibnamefont
  {Ivchenko}},\ }\bibfield  {title} {\enquote {\bibinfo {title} {Precession
  spin relaxation mechanism caused by frequent electron--electron
  collisions},}\ }\href@noop {} {\bibfield  {journal} {\bibinfo  {journal}
  {JETP Letters}\ }\textbf {\bibinfo {volume} {75}},\ \bibinfo {pages} {403}
  (\bibinfo {year} {2002})}\BibitemShut {NoStop}%
\bibitem [{\citenamefont {D'Amico}\ and\ \citenamefont
  {Vignale}(2003)}]{amico:045307}%
  \BibitemOpen
  \bibfield  {author} {\bibinfo {author} {\bibfnamefont {Irene}\ \bibnamefont
  {D'Amico}}\ and\ \bibinfo {author} {\bibfnamefont {Giovanni}\ \bibnamefont
  {Vignale}},\ }\bibfield  {title} {\enquote {\bibinfo {title} {Spin {C}oulomb
  drag in the two-dimensional electron liquid},}\ }\href
  {http://link.aps.org/abstract/PRB/v68/e045307} {\bibfield  {journal}
  {\bibinfo  {journal} {Phys. Rev. B}\ }\textbf {\bibinfo {volume} {68}},\
  \bibinfo {eid} {045307} (\bibinfo {year} {2003})}\BibitemShut {NoStop}%
\bibitem [{\citenamefont {Weng}\ and\ \citenamefont {Wu}(2003)}]{wu03prb}%
  \BibitemOpen
  \bibfield  {author} {\bibinfo {author} {\bibfnamefont {M.~Q.}\ \bibnamefont
  {Weng}}\ and\ \bibinfo {author} {\bibfnamefont {M.~W.}\ \bibnamefont {Wu}},\
  }\bibfield  {title} {\enquote {\bibinfo {title} {Spin dephasing in $n$-type
  $\mbox{GaAs}$ quantum wells},}\ }\href@noop {} {\bibfield  {journal}
  {\bibinfo  {journal} {Phys. Rev. B}\ }\textbf {\bibinfo {volume} {68}},\
  \bibinfo {pages} {75312} (\bibinfo {year} {2003})}\BibitemShut {NoStop}%
\bibitem [{\citenamefont {Glazov}\ and\ \citenamefont
  {Ivchenko}(2004)}]{glazov04a}%
  \BibitemOpen
  \bibfield  {author} {\bibinfo {author} {\bibfnamefont {M.~M.}\ \bibnamefont
  {Glazov}}\ and\ \bibinfo {author} {\bibfnamefont {E.~L.}\ \bibnamefont
  {Ivchenko}},\ }\bibfield  {title} {\enquote {\bibinfo {title} {Effect of
  electron-electron interaction on spin relaxation of charge carriers in
  semiconductors},}\ }\href@noop {} {\bibfield  {journal} {\bibinfo  {journal}
  {JETP}\ }\textbf {\bibinfo {volume} {99}},\ \bibinfo {pages} {1279} (\bibinfo
  {year} {2004})}\BibitemShut {NoStop}%
\bibitem [{\citenamefont {Alekseev}(2022)}]{alekseev:2}%
  \BibitemOpen
  \bibfield  {author} {\bibinfo {author} {\bibfnamefont {P.~S.}\ \bibnamefont
  {Alekseev}},\ }\bibfield  {title} {\enquote {\bibinfo {title} {Viscous flow
  of two-component electron fluid in magnetic field},}\ } {}
  {\bibfield  {journal} {\bibinfo  {journal} {Semiconductors}\ }\textbf
  {\bibinfo {volume} {56}},\ \bibinfo {pages} {650} (\bibinfo {year}
  {2022})}\BibitemShut {NoStop}%
\bibitem [{\citenamefont {Glazov}\ and\ \citenamefont
  {Golub}(2020{\natexlab{b}})}]{Glazov2020b}%
  \BibitemOpen
  \bibfield  {author} {\bibinfo {author} {\bibfnamefont {M.~M.}\ \bibnamefont
  {Glazov}}\ and\ \bibinfo {author} {\bibfnamefont {L.~E.}\ \bibnamefont
  {Golub}},\ }\bibfield  {title} {\enquote {\bibinfo {title} {{Skew Scattering
  and Side Jump Drive Exciton Valley Hall Effect in Two-Dimensional
  Crystals}},}\ }\href {\doibase 10.1103/PhysRevLett.125.157403} {\bibfield
  {journal} {\bibinfo  {journal} {Phys. Rev. Lett.}\ }\textbf {\bibinfo
  {volume} {125}},\ \bibinfo {pages} {157403} (\bibinfo {year}
  {2020}{\natexlab{b}})}\BibitemShut {NoStop}%
\end{thebibliography}
%

\end{document}